\documentstyle[preprint,aps]{revtex}
\input{psfig}
\begin{document}
\draft
\title{Measurements of the reactions $^{12}$C$(\nu_\mu$, $\mu^- )$
$^{12}$N$_{g.s.}$ and\\ $^{12}$C$(\nu_\mu$, $\mu^- )$ X}

\author{C. Athanassopoulos$^{12}$, L. B. Auerbach$^{12}$,
R. L. Burman$^7$, D. O. Caldwell$^3$, \\
E. Church$^{1}$, I. Cohen$^6$, B. D. Dieterle$^{10}$, J. B. Donahue$^7$,
A. M. Eisner$^4$,\\ A. Fazely$^{11}$,
F. J. Federspiel$^7$, G. T. Garvey$^7$, M. Gray$^3$, R. M. Gunasingha$^1$,\\
R. Imlay$^8$, K. Johnston$^{9}$,
H. J. Kim$^8$, W. C. Louis$^7$, R. Majkic$^{12}$,
\\ K. McIlhany$^{1}$, W. Metcalf$^8$, G. B. Mills$^7$,
R. A. Reeder$^{10}$, V. Sandberg$^7$, D. Smith$^5$,\\
I. Stancu$^{1}$, W. Strossman$^{1}$, R. Tayloe$^7$, G. J. VanDalen$^{1}$,
W. Vernon$^{2,4}$, N. Wadia$^8$,\\ 
J. Waltz$^5$, Y-X. Wang$^4$, D. H. White$^7$, 
D. Works$^{12}$, Y. Xiao$^{12}$, S. Yellin$^3$ \\
LSND Collaboration}
\address{$^{1}$University of California, Riverside, CA 92521}
\address{$^{2}$University of California, San Diego, CA 92093}
\address{$^3$University of California, Santa Barbara, CA 93106}
\address{$^4$University of California
Intercampus Institute for Research at Particle Accelerators,
Stanford, CA 94309}
\address{$^{5}$Embry Riddle Aeronautical University, Prescott, AZ 86301}
\address{$^6$Linfield College, McMinnville, OR 97128}
\address{$^7$Los Alamos National Laboratory, Los Alamos, NM 87545}
\address{$^8$Louisiana State University, Baton Rouge, LA 70803}
\address{$^{9}$Louisiana Tech University, Ruston, LA 71272}
\address{$^{10}$University of New Mexico, Albuquerque, NM 87131}
\address{$^{11}$Southern University, Baton Rouge, LA 70813}
\address{$^{12}$Temple University, Philadelphia, PA 19122}
 
\date{\today}
\maketitle

\begin{abstract}
Charged current scattering of $\nu_\mu$ on $^{12}$C has been studied using a
$\pi^+$ decay-in-flight $\nu_\mu$ beam at the Los Alamos Meson Physics
Facility.  A sample of 56.8 $\pm$ 9.6 events satisfying criteria for the exclusive
reaction $^{12}$C$(\nu_\mu$, $\mu^-$)$^{12}$N$_{g.s.}$ was obtained using the
large Liquid Scintillator Neutrino Detector (LSND).  The observed flux-averaged
cross section, $(6.6 \pm 1.0 \pm 1.0)\times 10^{-41} \rm{cm}^2$, agrees 
well with reliable theoretical expectations. 
A measurement was also obtained for the inclusive cross section to all accessible $^{12}$N
states, $^{12}$C($\nu_\mu$, $\mu^-$)X.  This flux-averaged cross section is
$(11.2\pm 0.3\pm 1.8)\times 10^{-40}$ cm$^2$, which is approximately half of that
given by a recent Continuum Random Phase Approximation (CRPA) calculation. 
\end{abstract}

\section{Introduction}

Low energy neutrino-nucleus cross sections are of interest because of their
application to nuclear structure studies and their importance as a base of 
information for low energy neutrino detectors. The cross sections contain
contributions from both axial vector and polar vector nuclear currents and thus
provide complementary information to that provided by electron-nucleus 
scattering, which is sensitive only to the nuclear polar vector currents.  

Thus far, data exist only for neutrino scattering on carbon.  Three 
experiments, E225 at LAMPF~[1],the KARMEN experiment at the ISIS facility of the
Rutherford Laboratory~[2] and LSND~[3], have measured both the exclusive 
reaction $^{12}\rm{C}(\nu_e, e^-)^{12}\rm{N}_{g.s.}$ and the inclusive 
cross section $^{12}\rm{C}(\nu_e, e^-)^{12}\rm{N}^*$ to all the accessible 
excited states of $^{12}$N. In these measurements the $\nu_e$ flux arises from 
$\mu^+$ decay at rest with E$_\nu <52.8$ MeV.  As a result of the low 
neutrino energy, transitions occur almost entirely to a few low lying states of
$^{12}$N, and 60 \% of the total cross section is to the $^{12}$N ground
state. The cross 
section for producing the $^{12}$N ground state can be predicted with an 
accuracy of $\approx$ 2\% by using model independent form factors that
can be reliably extracted from other measurements~[4].  All three  
experimental measurements of the $^{12}\rm{C}(\nu_e, e^-)^{12}\rm{N}_{g.s.}$ cross 
section agree well with
the expected value. Calculation of the inclusive cross section to the excited 
states of $^{12}$N is model dependent and is a less certain procedure.  The Fermi
Gas Model (FGM) is not reliable in this instance because the low neutrino 
energy leads to momentum transfers (q$<$100 MeV/c) much smaller than the Fermi
momentum (200 MeV/c) in carbon.  Thus extensive  modeling of the nuclear
dynamics is necessary.  A recent calculation~[5] that includes the 
particle-hole correlations in a continuum random phase approximation (CRPA) agrees
well with the $^{12}$C($\nu_e$, e$^-$)$^{12}$N$^*$ cross sections reported by 
the three experiments.

This paper expands on our earlier preliminary~[6,7] results on the 
charged current $\nu_\mu$ scattering
from carbon at somewhat higher neutrino energies using the flux of $\nu_\mu$ 
created by $\pi^+$ decay in flight.  The inclusive cross section is strongly 
energy and momentum transfer dependent.  Thus the flux-averaged cross section for the reaction
$^{12}$C($\nu_\mu$, $\mu^-$)$^{12}$N$^*$ is approximately 200 times larger than
the lower energy cross section for $^{12}$C($\nu_e$, e$^-$)$^{12}$N$^*$. 
In this case, a CRPA calculation yields a cross section approximately twice
the observed value~[8].  This apparent discrepancy, between the good agreement of 
the CRPA calculation and measurements for the decay-at-rest result and the 
factor of two discrepancy between a similar calculation and this measurement 
for the decay-in-flight cross section, has generated considerable theoretical
interest~[8]. 
A large-basis shell-model calculation, however, obtains a result that is
lower than the CRPA calculation for this reaction due to nuclear 
structure effects and may be consistent within errors with the observed
value~[9].
The only experiment previous to LSND had 
limited data and reported a cross section substantially above expectations and a 
muon energy spectrum much softer than expected~[10].  
These results are inconsistent with the LSND results~[6,7].

This paper presents a detailed study of $\nu_\mu$ carbon scattering based on 
data obtained in 1994 and 1995.  Both the exclusive reaction 
$^{12}$C($\nu_\mu$,$\mu^-$)$^{12}$N$_{g.s.}$ and the inclusive reaction 
$^{12}$C($\nu_\mu$,$\mu^-$)$^{12}$N are measured.  The calculated $\nu_\mu$ 
energy spectrum arising from $\pi^+$ decay-in-flight is shown in Fig. 1. 
Also shown is the energy spectrum of the much smaller $\bar{\nu}_\mu$ flux arising from
$\pi^-$ decay-in-flight.
Neutrinos with energies between muon production threshold (123.1 MeV) and 
approximately 280 MeV contribute to the cross section; therefore much higher 
nuclear excitation energies are possible for $\nu_\mu$C scattering than for 
the $\nu_e$C measurement discussed above. The measurement of the exclusive 
reaction, $^{12}$C($\nu_\mu$,$\mu^-$)$^{12}$N$_{g.s.}$,
provides a valuable check on the overall analysis procedure because
the cross section for this process can be reliably calculated in a mostly model 
independent fashion.

The analysis presented in this paper for events arising from
decay-in-flight neutrinos is also important because of its relevance to the
two searches for neutrino oscillations by LSND.  Evidence has been 
presented~[11] for $\bar{\nu}_\mu \rightarrow \bar{\nu}_e$ oscillations using
$\bar{\nu}_\mu$ from $\mu^+$ decay at rest.  The backgrounds to this 
measurement from decay-in-flight neutrinos are expected to be small, but it is
nevertheless important to measure these processes. LSND is also searching for 
$\nu_\mu \rightarrow \nu_e$ oscillations using the decay-in-flight $\nu_\mu$ 
beam by detecting electrons from the process $^{12}$C($\nu_e$,e$^-$)$^{12}$N. 
For this search quantitative knowledge is required both of the decay-in-flight 
neutrino beam and of neutrino-carbon cross sections.

\section{The Neutrino Source}

The data reported here were obtained in 1994 and 1995 at the Los Alamos
Meson Physics Facility (LAMPF) primarily using neutrinos produced at the A6 proton
beam stop. As discussed below some neutrinos are also produced at upstream 
targets A1 and A2.
  The neutrino source is described in detail elsewhere~[12].
This facility is now referred to as the Los Alamos Neutron Science Center (LANSCE).  The
beam stop consists of a 30-cm water target and a 50-cm pion decay region 
surrounded by steel shielding and
followed by a copper beam dump.  The high-intensity 800 MeV proton beam from
the linear accelerator generates a large pion flux from the water target. 
The fluxes of $\nu_\mu$ and $\bar{\nu}_\mu$ used for the measurements reported
here arise from the decay in flight (DIF) of $\pi^+$ and $\pi^-$.  For the
LAMPF proton beam and beam stop configuration $\pi^+$ production exceeds
$\pi^-$ production by a factor of approximately eight and even more for high
energy pions.  Approximately 3.4\% of the $\pi^+$ and 5\% of the $\pi^-$
decay in flight.  The LAMPF proton beam typically had a current of 800
$\mu$A and an energy of approximately 770 MeV at the A6 beam stop.  The
integrated beam current was 5904 C in 1994 and 7081 C in 1995.   Upstream
targets contributed 6\% to the DIF neutrino flux.  For the 1995 run, the water target
was removed for 32\% of the 7081 C of beam.  For this portion of the run
the DIF $\nu_\mu$ flux was reduced approximately 50\%.  The $\nu_\mu$ flux above
muon production threshold (123.1 MeV) and averaged over the LSND detector
was then $6.75\times 10^{11}$ cm$^{-2}$ for 1994 and $6.50\times 10^{11}$
cm$^{-2}$ for 1995.  The $\bar{\nu}_\mu$ flux above threshold (113.1 MeV)
for the process $\bar{\nu}_\mu +$ p $\rightarrow$ $\mu^+ +$ n was $6.79\times 10^{10}$
cm$^{-2}$ for 1994 and $6.69\times 10^{10}$ cm$^{-2}$ for 1995.

A detailed beam simulation program has been developed over the last
decade to describe the LAMPF beam dump which has been used as the neutrino
source for previous
experiments E31~[13], E225~[1] and E645~[14].  A calibration experiment,
E866~[15], measured the rate of stopped $\mu^+$ from a low-intensity
proton beam incident on an instrumented beam stop.  The rate of stopped
$\mu^+$ per incident proton was measured as a function of several variables
and used to fine tune the beam dump simulation program~[16].  This
simulation program can then be used to calculate the flux for any particular
beam dump configuration.  

The calibration experiment determined the decay-at-rest flux
to $\pm$ 7\% uncertainty for the proton energies and beam stop configurations used at
LAMPF.  There are greater uncertainties in the DIF fluxes.  Uncertainties in
the energy spectra of the $\pi^\pm$ which decay in flight lead to
uncertainties in both the magnitudes and shapes of the $\nu_\mu$ and
$\bar{\nu}_\mu$ energy spectra.  The resulting uncertainty in the DIF flux
for neutrinos above muon production threshold is estimated to produce
an uncertainty in the measured cross section of 15\%.

    We have performed a significant test of the beam simulation by
comparing data taken with two distinct beam dump configurations. As
discussed above, the water target was removed for 32\% of the 1995 data.
Per Coulomb of proton beam the beam simulation program predicts a 
$\nu_\mu$ flux above muon production threshold only 48\% as large with the
water target out as with the water target in. The predicted $\nu_\mu$ energy 
spectrum is, however, harder with the water target out, so that the predicted
rate for $^{12}$C$(\nu_\mu$, $\mu^- )$ X events is 61 $\pm$ 4\% as large 
and the mean detected energy
of these events is 15 $\pm$ 4\% larger. The systematic errors shown
include a conservative estimate of the effect of the uncertainty
 in the energy dependence of  the cross section. 
For comparison, the measured event rate
with the water target out is 57 $\pm$ 5\% of the rate with the water target
in. The mean event energy is 
13 $\pm$ 4\% larger with the water
target out.  The good agreement with expectations for both the event rate
and mean detected energy provides a valuable check of the beam simulation
program.

\section{The LSND Detector}

The detector is located 29.8 m downstream of the proton beam stop at an
angle of 12$^\circ$ to the proton beam.  Fig. 2 shows a side view of the
setup.  Approximately 2000 g/cm$^2$ of shielding above the detector
attenuates the hadronic component of cosmic rays to a low level. 
Enclosing the detector, except on the bottom, is a highly efficient liquid
scintillator veto shield which is essential to reduce contributions from the
cosmic ray muon background to a low level.  The detector is also well
shielded from the beam stop so that beam-associated neutrons are attenuated
to a negligible level.  Ref. 12 provides a detailed description of the
detector, veto and data acquisition system which we briefly review here.

The detector is a nearly cylindrical tank containing 167 tons of liquid
scintillator and viewed by 1220 uniformly spaced 8'' Hamamatsu
photomultiplier tubes (PMTs) covering
$\sim$25\% of the surface inside the tank wall.  The digitized time and
pulse height of each of these PMTs (and of each of the 292 veto shield PMTs)
are recorded when the deposited energy in the tank exceeds a threshold of
approximately 4 MeV electron-equivalent energy, and there are fewer than 4
PMT hits in the veto shield.  A veto, imposed for 15.2 $\mu$s following the
firing of $>$ 5 veto PMTs, substantially reduces $(10^{-3})$ the large
number of background events arising from the decay of cosmic ray muons that
stop in the detector.  Activity in the detector or veto shield during the
51.2 $\mu$s preceding a primary trigger is also recorded provided there are
$>$ 17 detector PMT hits or $>$ 5 veto PMT hits.  This activity information
is used in the analysis also to identify events arising from muon decay.  In
particular, in this analysis the activity information is used to identify
low energy $\mu^-$ from the reaction $\nu_\mu$ + $^{12}$C$\rightarrow \mu^-
+$ X.  For such events the e$^-$ from the subsequent decay $\mu^- \rightarrow$ e$^- +
\nu_\mu + \bar{\nu}_e$ provides the primary trigger.  It should also be
noted that the 15.2 $\mu$s veto applies only to the primary trigger and not
to the activities preceding a valid trigger.
   
Subsequent to a primary event trigger, data are recorded for 1 ms with a
greatly reduced threshold of 21 PMTs (approximately 0.7 MeV electron energy 
equivalent).  This low threshold is necessary to detect $\gamma$'s associated
with neutron capture, as described below.  The detector operates
without reference to the beam spill, but the state of the beam is recorded
with the event.  Approximately 93\% of the data is taken between beam
spills.  This allows an accurate measurement and subtraction of cosmic ray
background surviving the event selection criteria.

The detector medium consists of mineral oil (CH$_2$) in which is
dissolved a small concentration (0.031 g/l) of b-PBD~[17].  This mixture
allows the detection of both Cerenkov light and approximately isotropic 
scintillation light and produces
about 33 photoelectrons per MeV of electron energy deposited in the oil. 
The combination of the two sources of light provides direction information
and makes particle identification (PID) possible for relativistic particles. 
Identification of neutrons is accomplished through the detection of the 2.2
MeV $\gamma$ from neutron capture on free protons.  Stopping $\mu^-$ are
captured on $^{12}$C 8\% of the time in the LSND detector.  
The $\mu^\pm$ which decay
are readily identified as muons by the presence of subsequent spatially
correlated Michel electrons.

The veto shield encloses the detector on all sides except the bottom.  The
main veto shield~[18] consists of a 15-cm layer of liquid scintillator. 
Additional counters were placed below the veto shield after the 1993 run to
reduce cosmic ray background entering through the bottom support structure. 
These counters around the bottom support structure are referred to as bottom
counters.  A veto inefficiency $< 10^{-5}$ is achieved with this veto system
for incident charged particles.

\section{Analysis Techniques}

In the analysis presented in this paper we require a $\mu^\pm$
followed by a delayed coincidence with a decay e$^\pm$.  As a result of this
coincidence requirement a clean beam excess sample of events can be obtained
with relatively loose selection criteria.  Furthermore, it is easy to verify
that the events in this sample arise from muon decay since the muon lifetime
and the decay electron energy spectrum are well known, and the response of
the LSND detector to electrons from muon decay has been well determined from
a large, clean sample of electrons from decays of stopping cosmic ray muons.  

Each event is reconstructed using the hit time and pulse height of all hit
PMTs in the detector.  The present analysis relies on the
reconstructed energy, position and particle ID parameter,
$\chi_{tot}$~[12].  
The parameter $\chi_{tot}$ is used to distinguish electrons
from interactions of cosmic ray neutrons in the
detector and will be explained below.

Fortunately, it is possible to measure the response of the detector to
electrons and neutrons in the energy range of interest for this analysis. 
  The response of
the detector to electrons was determined from a large, essentially pure
sample of electrons (and positrons) from the decay of stopped cosmic ray
$\mu^\pm$ in the detector.  The known energy spectrum for electrons from muon
decay was used to determine the absolute energy calibration including its
small variation over the volume of the detector.  The energy resolution was
determined from the shape of the electron energy spectrum as shown in Fig.
3 and was found to be 6.6\% at the 52.8 MeV endpoint.  We also make use of a
detailed Monte Carlo simulation, LSNDMC~[19], which
was written to simulate events in the detector using GEANT.  The position
resolution obtained from the LSNDMC simulation is approximately 30 cm for 
a 20 MeV electron.

For relativistic electrons in the LSND detector approximately 70\% of the
photoelectrons arise from direct or re-radiated Cerenkov light and only 30\%
from scintillator light.  For muons, the threshold kinetic energy for
Cerenkov radiation in the LSND detector is 39 MeV.  For the sample of muons
analyzed in this paper only about half are above Cerenkov threshold
and none fully relativistic.  As a result, the light output per MeV of energy
loss for the muons is significantly less than that for relativistic
electrons.  There is no calibration sample available of low energy muons
with known energies.  Thus we rely on the Monte Carlo simulation LSNDMC for
muons.  We discuss the muon energy scale further in sections VI and VII when
we compare observed and expected energy distributions.  

There are no tracking devices in the LSND detector, and thus event positions
must be determined solely from the PMT information.  The reconstruction
process determines an event position by minimizing a function
$\mathcal{X}$$_r$ which is based on the time of each PMT hit corrected for
the travel time of light from the assumed event position to the PMT~[12].
This reconstruction procedure was found to systematically shift event
positions away from the center of the detector and thus effectively reduces
the fiducial volume~[12], as discussed below.  In the analysis presented in this paper a
fiducial cut is imposed on the electron by requiring D $>$ 35 cm, where D is
the distance between the reconstructed electron position and the surface
tangent to the faces of the PMTs.

The effect of the reconstruction bias on the fiducial acceptance was
determined from the analysis of a sample of stopping muon events for which
both the muon and the subsequent decay electron were detected.  No fiducial
cut was imposed on either the muon or the electron so that essentially all
muons which stopped in the scintillator and decayed were included.  For
comparison a sample of simulated stopping muon events was generated using
LSNDMC.  The observed and generated distributions of the distance D were
compared for electrons satisfying a minimum energy requirement.  The
observed distribution was found to be shifted outward relative to the
generated distribution.  Several independent analyses of this type yielded the
acceptance factor of $0.85\pm 0.05$ for D $>$ 35 cm due to the
reconstruction bias.  There is independent support for this conclusion.  A
new reconstruction procedure has been developed which relies primarily on
PMT pulse height rather than timing information and is not expected to have
a significant bias.  Comparison of vertex positions obtained with the new and the standard
reconstruction procedures indicate an outward shift in good agreement
with that obtained from the stopping muon analysis.

The particle identification procedure is designed to separate particles with
velocities well above Cerenkov threshold from particles below Cerenkov
threshold by making use of the four parameters defined in Ref. 12.
Briefly, $\mathcal{X}$$_r$ and $\mathcal{X}$$_a$ are the quantities
minimized for the determination of the event position and direction,
$\mathcal{X}$$_t$ is the fraction of PMT hits that occur more than 12 ns
after the fitted event time and $\chi_{tot}$ is proportional to the
product of $\mathcal{X}$$_r$, $\mathcal{X}$$_a$ and $\mathcal{X}$$_t$.  For
the present analysis we use only $\chi_{tot}$ and impose a
requirement only on the electron candidate, not on the preceeding muon
candidate.  Figure 4 shows the $\chi_{tot}$ distributions for
electrons from stopping $\mu$ decay and for cosmic ray neutrons with
electron-equivalent energies in the 16 $<$ E$_e <$ 60 MeV range.  
For a neutron E$_e$ is the equivalent electron energy corresponding to the
observed total charge.  In the present analysis a relatively loose
$\chi_{tot}$ requirement reduces the neutron background to a
negligible level.

The presence of a neutron can be established from the neutron capture reaction n
+ p $\rightarrow$ d + $\gamma$.  The mean capture time in the LSND detector
is expected to be 186 $\mu$s, essentially independent of the initial neutron
energy.  Three variables are used to identify a capture $\gamma$ correlated
with a neutron in the primary event:  the number of PMT hits for the
$\gamma$, the distance of the $\gamma$ from the primary event and the time
of the $\gamma$ from the primary event.  Fig. 5 shows the distributions of
these variables for correlated $\gamma$'s and for uncorrelated (accidental)
$\gamma$'s.  A likelihood technique, discussed in Ref. 11, has been
developed to separate the correlated component due to neutrons from the
uncorrelated component.  An approximate likelihood ratio $R\equiv
\mathcal{L}$$_{cor}$/$\mathcal{L}$$_{uncor}$ is calculated for each event
from the three measured variables.  If there is no $\gamma$ within 1 ms and
2.5 m from the primary event then R = 0 for the event.  The expected
distributions of R are shown in Fig. 6 for a correlated sample (every
event has one neutron) and for an uncorrelated sample (no event has a
neutron).  The correlated R distribution was found to be almost independent
of event position within the fiducial volume~[11].  The accidental gamma
rate is higher near the bottom front corner of the detector than elsewhere,
but the shape of the uncorrelated R distribution has little position
dependence.  Also shown in Fig. 6 is the measured R distribution~[3] for a
clean sample of over 500 events from the reactions $^{12}$C($\nu_e$,
e$^-$)$^{12}$N$_{g.s.}$.  Such events have no associated neutrons and thus
this sample provides a useful check of the uncorrelated R distribution.  The
agreement with the distribution for uncorrelated $\gamma$'s is excellent.  
In the present paper
we use the $\gamma$ analysis to determine the fraction of the events in the
DIF sample that are accompanied by neutrons.
The measured R distribution is fit to a mixture of the
correlated and uncorrelated distributions shown in Fig. 6, and the fraction
of events with neutrons is obtained.

Beam-off data taken between beam spills play a crucial role in the analysis
of this experiment.  Most event selection criteria are designed to reduce
the background due to cosmic rays while retaining high acceptance for the neutrino
process of interest.  The cosmic ray background which remains after all
selection criteria have been applied is well measured with the beam-off data
and subtracted using the duty ratio, the ratio of beam-on time to beam-off
time.  This ratio was 0.080 for 1994 and 0.060 for 1995.  The smaller duty
ratio in 1995 arose from changes in LAMPF beam operations, especially a
reduction in the number of proton beam spills per second at the A6 beam
dump.

\section{Event Selection}

The analysis is designed to select the $\mu^-$ from the reaction $\nu_\mu~+~
^{12}$C $\rightarrow \mu^- +$ X and the subsequent electron from the decay $\mu^-
\rightarrow$ e$^- + \bar{\nu}_e + \nu_\mu$.  In the LSND detector medium 
92\% of the stopped
$\mu^-$ decay and 8\% are captured.  The $\mu^-$ and other particles arising
from the charge-changing neutrino interaction produce light that causes an average of 250
PMTs to fire.  The detector signal, Q$_\mu$, measured in photoelectrons
arises mostly from the $\mu^-$ but includes contributions from other
particles produced in the reaction such as protons and gammas.

Table I shows the selection criteria and corresponding efficiencies for the
muon and electron for events in which there are more than 100 PMT hits at
the time of the $\mu^-$.  Slightly tighter criteria, discussed below, are
used for the $\sim$10\% of the events with fewer than 100 PMT hits.  These
two samples are referred to as ``high energy $\mu$'' and ``low energy $\mu$'',
 respectively.
For events in the decay-in-flight sample the event position is best
determined from the reconstructed electron position rather than the
reconstructed muon position, especially for events with low energy muons.
Therefore, the
fiducial selection is imposed primarily on the electron.  The reconstructed
electron position is required to be a distance D $>$ 35 cm from the surface
tangent to the faces of the PMTs.  
There are 3.65 $\times$ 10$^{30}$ $^{12}$C
nuclei within this fiducial volume.  The muon is required to
reconstruct only inside the region D $>$ 0 cm.  A lower limit on the electron
energy of 16.0 MeV eliminates the large  background from $^{12}$B
beta decay created by the capture of cosmic ray $\mu^-$ on $^{12}$C.  Fig. 7
shows the observed electron energy distribtuion 
compared with that obtained from a large clean sample of Michel electrons from
decays of stopping cosmic ray muons.  The distribution of the time,
$\Delta$t$_{\mu e}$,
between the muon and electron candidates, shown in Fig. 8, agrees well
with the 2.03 $\mu$s $\mu^-$ lifetime in mineral oil.  The best fit, also 
shown, corresponds to a lifetime $1.98 \pm 0.06 \mu$s.
The requirement
$\Delta$t$_{\mu e}$ $\geq$ 0.7 $\mu$s is imposed to insure that the $\mu$ and e are
clearly separated in the trigger and in the readout of the data.  The
excellent agreement with expectations in Fig. 7 and 8 clearly show that
the events arise from muon decay.  
There is an 8\% loss of events due to $\mu^-$ capture in the detector medium.
Fig. 9 shows the spatial separation,
$\Delta$r, between the reconstructed muon and electron positions.  A loose
requirement, $\Delta$r $<$ 1.5 m, is imposed to minimize the background from
accidental $\mu$,e correlations while retaining high acceptance.

Many of the selection criteria are designed to reduce the cosmic ray
background, especially that arising from the decay of cosmic ray muons which
stop in the detector.  Events are required to have fewer than 4 PMT hits in
the veto at both the muon time and the electron time.  The detector PMT faces
are 25 cm inside the tank and thus stopping cosmic ray muons must traverse
at least 60 cm of scintillator to reach the fiducial volume.  As a result
these muons typically produce a large detector signal.  The
requirement Q$_\mu < 2000$ pe, where Q$_\mu$ is the detector signal
at the muon time measured
in photoelectrons, eliminates most such background events with almost no
loss of acceptance for muons arising from neutrino interactions.

Frequently, in addition to the candidate muon which satisfies the criteria
in Table I, there are one or more other activities prior to the electron. 
If an activity is due to a stopping muon, that muon could be the parent of
the observed electron.  Therefore an event is rejected if, in the 35 $\mu$s
interval prior to the electron, there is an activity with Q $>$ 3000 pe or
an activity with $>$ 4 PMT hits in the veto and $>$ 100 PMT hits in the
detector.  This results in a 7\% loss of efficiency for neutrino events. 
The efficiency for the past activity criteria shown in Table I also includes the
effect of the veto that is applied for 15.2 $\mu$s following any event with
$>$ 5 PMT hits in the veto shield.

The acceptances for the past activity and in-time veto cuts are obtained by
applying these cuts to a large sample of events triggered with the laser
used for detector calibration.  These laser events are spread uniformly
through the run and thus average over the small variation in run conditions. 

Only a loose particle ID requirement, $\mathcal{X}$$_{tot} < 1.0$, was
imposed on the electron and none on the muon.
A sample of Michel electrons (electrons from the decay
of stopped $\mu^\pm$) was analyzed to obtain the acceptance of electrons for
the $\chi_{tot}$ particle identification cut as shown in Fig. 4.

For events in which fewer than 100 PMT hits occur at the muon time (low
energy
$\mu$ events) tighter selection criteria are needed to keep the beam-off
background small.  These tighter criteria are: (1) electron
particle ID ($\mathcal{X}$$_{tot} < 0.8$ instead of 1.0), (2) muon decay
time ($\Delta$t$_{\mu e}$ $< 6.6 \mu$s instead of 9.0 $\mu$s) and
(3) a tighter past
activity cut.  As a result, the efficiency for this inclusive 
``low energy $\mu$'' sample is only
67 $\pm$ 1\% of the ``high energy $\mu$'' sample efficiency.  This includes the small
loss of acceptance for muons below the 18 PMT detection threshold.  

The Monte Carlo simulation LSNDMC was used to obtain the PMT hits
distributions expected from the various processes that contribute to the
inclusive sample and to the exclusive sample with an identified beta decay.  
Fig. 10 compares the observed and expected
distributions of PMT hits for inclusive events.  There is excellent agreement,
and thus we expect that the simulation provides a reliable estimate of the
fraction of ``low energy $\mu$'' events below 18 PMT hits (roughly 4 MeV).  For the inclusive sample
(exclusive sample) we find that only 6\% (18\%) of the ``low energy $\mu$'' events have
fewer than 18 PMT hits. The overall efficiency for the inclusive (exclusive)
``low energy $\mu$'' sample  is 67\% (61\%) of the efficiency for the 
``high energy $\mu$'' sample.

For analysis of the exclusive process $^{12}$C($\nu_\mu$, $\mu^-
)^{12}$N$_{g.s.}$ we also require detection of the e$^+$ from the beta decay
of $^{12}$N$_{g.s.}$.  Therefore, for these events three particles are
detected: the muon, the decay electron and the positron from the beta
decay of $^{12}$N$_{g.s.}$.
Table II gives the selection criteria and efficiencies
for the $^{12}$N beta decay positron.  These are the same criteria~[3] used
previously in an analysis of a sample of over 500 events from the analogous
process $^{12}$C($\nu_e$, e$^- )^{12}$N$_{g.s.}$.  The beta decay has a mean
lifetime of 15.9 ms and maximum positron kinetic energy of 16.3 MeV.
Fig.~11 shows the observed beta decay time distribution compared with the
expected 15.9 ms lifetime.  Fig. 12 shows the distance between the
reconstructed electron and positron positions for the beam-excess sample. 
For comparison, the distribution observed for the $^{12}\rm{C}(\nu_e, e^-)^{12}\rm{N}_{g.s.}$
sample is shown by the solid line.
A cut was applied at 100 cm resulting in an acceptance of (96 $\pm$ 2)\%.  The
positron is required to be spatially correlated with the electron rather
than the muon because the position of the electron in general is better
determined.  
Following a muon produced by a neutrino interaction, an uncorrelated
particle, such as the positron from $^{12}$B beta decay, will occasionally
satisfy all the positron criteria including the requirements of time (45 ms)
and spatial (1 m) correlation with the electron.  The probability of such an
accidental coincidence can be precisely measured from the Michel electron
sample.  The background from this source is also shown in Figs. 11 and 12. 
The efficiency of 81.5\% caused by the 15.2 $\mu$s veto and the trigger dead
time of 3\% are the same as for the electron.  Positrons with 4 or more
in-time veto hits or any bottom veto coincidence are rejected.  The Monte
Carlo was used to generate expected distributions for the positron energy. 
There was a trigger requirement of 100 PMT hits for 1994 and 75 PMT hits for
1995.  The positron was required to have an energy less than 18 MeV. 
Fig. 13 compares the observed and expected positron energy
distributions.  Fig. 14 compares the observed and expected energy
distributions of the electron from the muon decay, and Fig. 15 compares the
observed and expected distributions of muon decay time.

Table III shows the numbers of beam-on, beam-off and beam-excess events
satisfying the ``high energy $\mu$'' and ``low energy $\mu$'' selection criteria.  The ``low energy $\mu$''
events are given a weight to compensate for the lower efficiency
for this sample.  This allows the relatively small ``low energy $\mu$'' sample
to be combined with the ``high energy $\mu$'' sample for the rest of the analysis.
The sample of
inclusive $\mu$-e events is used for the analysis of the reaction $\nu_\mu +
^{12}$C $\rightarrow \mu^- +$ X, while the exclusive sample of events with an
identified beta decay is used for analysis of the reaction $\nu_\mu +
^{12}$C $\rightarrow \mu^- + ^{12}$N$_{g.s.}$.

\section{The Transition to the $^{12}$N Ground State}

The reaction $\nu_\mu +
^{12}$C $\rightarrow ^{12}$N$_{g.s.} + \mu^-$ is identified by the
detection of the $\mu^-$, the e$^-$ from the decay $\mu^- \rightarrow$ e$^-
+ \nu_\mu + \bar{\nu}_e$ and the positron from the $\beta$ decay of the
$^{12}$N$_{g.s.}$.  This three-fold delayed coincidence requirement provides a
distinctive event signature.  Excited states of $^{12}$N decay
by prompt proton emission and thus do not feed down to the $^{12}$N ground
state or contribute to the delayed coincidence rate.  The form factors
required to calculate the cross section are well known from a variety of
previous measurements~[4].  This cross section and the known $\nu_\mu$ flux are
used to obtain the expected muon kinetic energy spectrum which is compared
with the data in Fig. 16.

As stated in Section IV the energy calibration for muons (the conversion
from photoelectrons to MeV) is obtained from the Monte Carlo simulation
LSNDMC.  
For this ground state reaction, the expected muon energy distribution should be 
very reliable.  Thus the good agreement seen in Fig. 16 provides confirmation for the
muon energy calibration within the limited statistics.

There are two sources of background.  The largest arises from the accidental
coincidence of a positron candidate with an event from the inclusive sample
of neutrino-induced muons.  The probability of an uncorrelated particle
satisfying all the positron criteria, including the requirements of time (45
ms) and spatial correlation (1 m) with the electron, can be precisely measured
from a large Michel electron sample.  The probability was 0.57\% for 1994
and 0.84\% for 1995.  The probability is higher in 1995 because a lower PMT
threshold was required that year for the positron.  The second background arises from the
process $^{12}$C($\bar{\nu}_\mu$, $\mu^+$)$^{12}$B$_{g.s.}$, where we detect
the e$^-$ from the beta decay of the $^{12}$B ground state~[20].
  This background
is small primarily because the flux of high energy $\bar{\nu}_\mu$ is
approximately a factor of ten lower than the corresponding $\nu_\mu$ flux and
because the $^{12}$B$_{g.s.}$ lifetime is longer than the $^{12}$N$_{g.s.}$ lifetime.

Table IV shows the number of beam excess events, the number of
background events, the ``high energy $\mu$'' efficiency, the neutrino flux for E$_\nu
>$ 123.1 MeV and the cross section averaged over the flux.  The predicted
flux-averaged cross section shown in Table IV was calculated for the flux
shape for the 1994 LSND beam dump configuration and not for the average of
the two years of data.  Therefore, the measured flux-averaged cross section
in Table IV has been adjusted slightly so that it also corresponds to the
1994 configuration.  The flux-averaged cross section is
\[ <\sigma > = (6.6\pm 1.0\pm 1.0)\times 10^{-41} \rm{cm}^2 , \]
where the first error is statistical and the second systematic.  The
two dominant sources of systematic error are the neutrino flux (15\%)
discussed in section II and the effective fiducial volume (6\%) discussed in
section IV.  The measured cross section is in very good agreement with the
predicted cross section of $6.4 \times 10^{-41}$ cm$^2$~[20].  There is
very little uncertainty in this predicted cross section for the exclusive
process, as it is determined from measured form factors in related electro-weak
processes.  If we assume that it is correct, we can use our measurement to
determine the $\nu_\mu$ flux instead of the cross section.
This yields a value for the $\nu_\mu$
flux above the muon production threshold that is $(105 \pm 16)$\% of the
value calculated using the beam Monte Carlo, if we assume the shape of the
$\nu_\mu$ flux is correctly given by the Monte Carlo.  This excellent agreement 
provides a valuable confirmation of our understanding of the flux 
from the neutrino source. 

For this reaction to the $^{12}$N ground state it is also straightforward to
measure the energy dependence of the cross section.  The recoil energy of
the nucleus is small and to a very good approximation, E$_\nu =$
m$_\mu$c$^2 +$ T$_\mu +17.7$ MeV,  where m$_\mu$ is the muon mass,
T$_\mu$ the muon kinetic energy and 17 .7 MeV arises from the Q value of the
reaction and the nuclear recoil.  Fig. 17 compares the measured cross section as a
function of E$_\nu$ with four theoretical calculations obtained from Ref.
20.  The agreement is excellent within the limited
statistics.

\section{The Inclusive Reaction}

Most of the inclusive beam-excess events arise from the reaction
$^{12}$C($\nu_\mu$, $\mu^-$)X, but approximately 10\% are due to other
sources.  Table V shows the number of beam-excess events, the calculated
backgrounds, the ``high energy $\mu$'' efficiency, $\nu_\mu$ flux and 
the flux-averaged cross
section for this process.  The backgrounds arising from the $\bar{\nu}_\mu$
component of the decay-in-flight beam are small, primarily because the high
energy $\bar{\nu}_\mu$ flux is approximately a factor of ten lower than the
corresponding $\nu_\mu$ flux.  The largest background arises from the
process $\bar{\nu}_\mu +$ p $\rightarrow \mu^+ +$ n.  The cross
section is well known and the uncertainty in this process is mainly due to
the 15\% uncertainty in the $\bar{\nu}_\mu$ flux.  A much smaller but less
well understood background arises from the process $^{12}$C($\bar{\nu}_\mu$,
$\mu^+$)X.  Plausibly, as observed for the process $^{12}$C($\nu_\mu$,
$\mu^-$)X, the cross section might be expected to be approximately 60\% of
that given by a recent CRPA calculation~[8].  We use this reduced cross section
in calculating this background but assign a large error to reflect the
uncertainty in the cross section.  An even smaller background arises from
the 1.1\% $^{13}$C component of the scintillator.  For the process
$^{13}$C($\nu_\mu$, $\mu^-$)X we use a Fermi Gas Model calculation and assign
a 50\% uncertainty.

The measured flux-averaged cross section for the inclusive
reaction $^{12}$C($\nu_\mu$, $\mu^-$)X is 
\[ <\sigma > = (11.2\pm 0.3\pm 1.8)\times 10^{-40} \rm{cm}^2 , \]  
where the first error is statistical and the
second  systematic.
The mean energy of the neutrino flux above
threshold is 156 MeV.
The systematic error is due almost entirely to the
uncertainty in the $\nu_\mu$ flux.  
The inputs to the neutrino beam Monte
Carlo program were varied within the estimated uncertainties.  The resulting
variation in both the magnitude and the shape of the $\nu_\mu$ flux above
muon production threshold results in a 15\% uncertainty in the inclusive
cross section.  As discussed in section VI, the flux-averaged cross section
has been adjusted so that it corresponds to the flux shape from the 1994
beam dump configuration and not the average of the two years of data.
This was done to permit direct comparison with the CRPA calculation of Ref.~[8]
which was performed for the 1994 $\nu_\mu$ flux. Ref.~[8] predicts a 
flux-averaged cross section of $20.5 \times 10^{-40} \rm{cm}^2$ which is 
significantly higher than that measured.
The measured cross section reported in this paper is 2$\sigma$ higher than
that originally reported by LSND~[6].  A substantial part of the 
increase arises from a better understanding of the loss of acceptance due to 
the spatial reconstruction program shifting events outward as discussed in 
section IV. 

The spatial distribution of the beam-excess electrons is shown in Fig. 18.
There is a clear enhancement of events at high x and high y due to the
variation of the $\nu_\mu$ flux over the detector.  The good agreement
with expectation shows that this spatial distribution is well modeled
by the beam simulation program.

For the reaction $^{12}$C($\nu_\mu$, $\mu^-$)X, the detector signal, Q$_\mu$,
measured in photoelectrons, arises mostly from the $\mu^-$ but includes
contributions from other particles in the reaction such as protons and gammas.  The muon
kinetic
energy distribution can be obtained from the $Q_\mu$ distribution by subtracting
the contributions of these other particles. The average contributions from proton
and $\gamma$'s are estimated to be 9 MeV and 2.9 MeV, respectively~[8].
We used the calculation of Ref. 8
to determine proton and $\gamma$ energy distributions and LSNDMC to determine the number
of photoelectrons produced. Protons produce less scintillation light than electrons due to 
saturation effects.  The uncertainty in the saturation effect is the primary source
of uncertainty in the muon and proton energy determination.
The average contribution to Q$_{\mu}$ from particles other 
than the muon is approximately
20\% using the CRPA calculation.  The information available in Ref. 8
permits us to subtract the protons and $\gamma$ contributions with a
procedure that is only correct on average.  The resulting E$_\mu$ distribution
should, however, be fairly reliable since the mean correction to Q$_\mu$ is
only 20\%.
Fig. 19 compares the observed distribution of E$_\mu$ with the shape
expected
from a recent CRPA calculation which has been normalized to the data.  
 There is fair agreement.  However, given the
uncertainties in the shape of the $\nu_\mu$ energy spectrum, 
in  the modeling of the energy
from nuclear breakup and in the muon and proton energy calibration, 
we do not try to extract any information on
the energy dependence of the cross section for the reaction
$^{12}$C($\nu_\mu$,$\mu^-$)X.

Further information on the inclusive sample can be obtained by measuring the
fraction of the events with an associated neutron.  The presence of a
neutron is established by detection of the gamma ray from the neutron's
capture on a proton in the detector via the reaction n + p
$\rightarrow$ d + $\gamma$.  The $\gamma$'s are detected using the procedure discussed in section IV.
The distribution of the likelihood ratio R for correlated $\gamma$'s from
neutron capture is very different from that for uncorrelated (accidental)
$\gamma$'s.  The measured R distribution for the inclusive sample, shown in
Fig. 20, was fit to a mixture of the two possible gamma sources to determine
the fraction of events with associated neutrons.  The best fit, also shown in the
figure, corresponds to a fraction of events with a neutron of
(10.8 $\pm$ 1.8)\%, where the error includes systematic uncertainties.
 
The two largest backgrounds shown in Table V arise from the $\bar{\nu}_\mu$
component of the beam, and almost all of these events should have an
associated neutron. In contrast, most of the events arising from $\nu_\mu$
interactions will not have an associated neutron.  A CRPA calculation
predicts that 79\% of the events from the reaction
$^{12}$C($\bar{\nu}_\mu$,$\mu^+$)X will have an associated neutron compared to only
5.9\% for the reaction $^{12}$C($\nu_\mu$,$\mu^-$)X~[8].

Table VI shows the measured component with an associated neutron for the
beam excess sample, the calculated backgrounds from $\bar{\nu}_\mu$
interactions and the resulting numbers for the $\nu_\mu$ carbon sample.
The percentage of events with neutrons for the $\nu_\mu$ carbon sample,
(1.9 $\pm$ 2.6)\%, is lower than but consistent with the CRPA prediction of
5.9\%.  The
observed number of events with neutrons also rules out a $\bar{\nu}_\mu$ flux much
bigger than that calculated by the beam Monte Carlo. 

\section{Conclusions}

The exclusive process $^{12}$C($\nu_\mu$,$\mu^-$)$^{12}$N$_{g.s.}$ has been
measured with a clean sample of 56.8 $\pm$ 9.6 events for
which the $\mu^-$, the decay e$^-$ and
the e$^+$ from the beta decay of the $^{12}$N$_{g.s.}$ are detected.  For this
process the cross section calculations are very reliable. 
The flux-averaged cross section is measured to be
(6.6$\pm$1.0$\pm$1.0)$\times 10^{-41}$ cm$^2$ in good agreement with
theoretical expectations.  From comparison of this cross section with the 
cross section for the inclusive process $^{12}$C($\nu_\mu$,$\mu^-$)X we obtain
a flux-averaged branching ratio of $(5.9 \pm 0.9 \pm 0.6)\%$

The inclusive process $^{12}$C($\nu_\mu$,$\mu^-$)X has also been measured. 
There are model dependent uncertainties in the calculated cross section for this
process that are not present for the $^{12}$N$_{g.s.}$ cross section. 
The flux-averaged cross
section is found to be (11.2$\pm$0.3$\pm$1.8)$\times 10^{-40}$ cm$^2$, about
55\% of a recent CRPA calculation~[8].
The mean energy of the neutrino flux above
threshold is 156 MeV.
  The measured distribution of the muon
energy (including contributions from other particles such as  protons and
gammas) agrees within errors with the CRPA calculation~[8].  The fraction of
events with associated neutrons was  measured to be (1.9 $\pm$ 2.6)\%.
This is lower than, but consistent with, the CRPA calculation of 5.9\%.

As discussed above, there has been considerable interest in the fact that
our observed cross section for the inclusive reaction $\nu_\mu + ^{12}$C
$\rightarrow \mu^- +$ X is only $0.55 \pm 0.09$ of the result obtained in a
sophisticated CRPA calculation~[8] for the same process.  Such CRPA
calculations have to be tuned to fit the cross sections to final bound
states, but without further adjustments to continuum final states.  The CRPA
calculation
also reproduces the inclusive cross section (to well within the 17\%
experimental error) for the process $\nu_e + ^{12}$C $\rightarrow$ e$^- +$
$^{12}$N$^*$, where the $\nu_e$ arise from $\mu^+$ decay at rest 
(E$_\nu <$ 52.8 MeV)~[3]. 
The situation has been discussed in some detail in Ref. 8 but we wish to
make some further observations.  First, the agreement between the CRPA
calculation of $\nu_\mu + ^{12}$C $\rightarrow \mu^- +$
X and the newer data presented here is better, $55\pm 9\%$ rather than
43$\pm$9\%.  A possible reason that the calculation agrees with $\mu^-$
capture and the low energy $\nu_e$ inclusive scattering but fails in the
case of the higher energy $\nu_\mu$ is that both the muon capture and the
$\nu_e$ processes involve only low partial waves $l = 0, 1$. 
The momentum
transfers for the $\nu_\mu$ reaction (100-400 MeV/c) are much larger than
for the lower energy $\nu_e$ and muon capture processes ($<$100 MeV/c). 
Thus the former reaction can involve partial waves through $l = 3, 4$, 
values not
probed by the lower energy processes.  It is necessary to point out that no
conclusion can be drawn from our results on the ratio of $\nu_\mu$ to
$\nu_e$ cross sections as the measurements involved very different neutrino
energies and momentum transfers.  However, it appears that factor of two
discrepancies may exist between sophisticated calculations and measured
inclusive yields.  This makes it difficult to have confidence that for the
lower energy atmospheric neutrino data~[21,22], 
which is only slightly above the LSND energy region, 
the observed lepton rates can 
be reliably converted into absolute neutrino fluxes.

\paragraph*{Acknowledgments}

The authors gratefully acknowledge the support of Peter Barnes,
Cyrus Hoffman, and John McClelland. 
This work was conducted under the auspices of the US Department of Energy,
supported in part by funds provided by the University of California for
the conduct of discretionary research by Los Alamos National Laboratory.
This work is also supported by the National Science Foundation.
We are particularly grateful for the extra effort that was made by these
organizations to provide funds for running the accelerator at the end of
the data taking period in 1995.

\clearpage

\begin{table}
\caption{The muon and electron selection criteria and corresponding
efficiencies for events with more than 100 PMT hits at the muon time.}
\label{I}
\begin{tabular}{lcl}
Quantity & Criteria & Efficiency\\ \hline
Fiducial Volume e & D$>$35.0 cm & 0.850$\pm$0.050\\
Fiducial Volume $\mu$ & D$>$0 & 0.950$\pm$0.005\\
Electron Energy & 16$<$E$_e$$<$60 MeV & 0.890$\pm$0.005\\
Muon Charge & Q$<$2000 pe & 0.990$\pm$0.010\\
Electron Particle ID & $\chi_{tot}$$<$1.0 & 0.976$\pm$0.005\\
Intime Veto $\mu$, e & $<$4 PMTs & 0.984$\pm$0.007\\
Past Activity & $\Delta$t$_p$$>$35 $\mu$s & 0.750$\pm$0.010\\
$\mu$ Decay Time & 0.7$<$t$<$9.0 $\mu$s & 0.687$\pm$0.005\\
Not $\mu^-$ Capture & & 0.922$\pm$0.005\\
Spatial Correlation & $\Delta$r$<$1.5 m & 0.993$\pm$0.002\\
DAQ Dead Time & & 0.970$\pm$0.010\\ \hline
Total & & 0.313$\pm$0.020\\ 
\end{tabular}
\end{table}

\begin{table}
\caption{Beta decay e$^+$ selection criteria and corresponding efficiencies
for the reaction $^{12}$C($\nu_\mu$, $\mu^-$)$^{12}$N$_{g.s.}$.}
\label{II}
\begin{tabular}{lcl}
Quantity & Criteria & Efficiency\\ \hline
$\beta$ Decay Time & 52 $\mu$s$<$t$<$45 ms & 0.938$\pm$0.002\\
Spatial Correlation & $\Delta$r$<$1 m & 0.964$\pm$0.020\\
PMT Threshold & $>$100 for 1994, $>$75 for 1995 & 0.823$\pm$0.015\\
Fiducial Volume & D$>$0 cm & 0.972$\pm$0.010\\
Trigger Veto & $>$15.2 $\mu$s & 0.815$\pm$0.005\\
Intime Veto & $<$4 PMTs & 0.992$\pm$0.001\\
DAQ Dead Time & & 0.970$\pm$0.010\\ \hline
Total & & 0.568$\pm$0.017\\ 
\end{tabular}
\end{table}

\begin{table}
\caption{Inclusive events and events with an identified beta decay. 
Events are classified as ``high energy $\mu$'' (``low energy $\mu$'') if there are more than (less than)
100 PMT hits at the time of the muon.  The ``low energy $\mu$'' events are given a weight
to compensate for the reduced efficiency for the ``low energy $\mu$'' sample.}
\label{III}
\begin{tabular}{|l|c|c|c|c|}
\multicolumn{1}{c}{} & \multicolumn{2}{c}{Inclusive Events} & \multicolumn{2}{c}{Events with beta
decay}\\ \hline
& ``high energy $\mu$'' & ``low energy $\mu$'' & ``high energy $\mu$'' & ``low energy $\mu$''\\
& $>$100 PMTs & $<$100 PMTs & $>$100 PMTs & $<$100 PMTs\\ \hline
Beam-on & 1755 & 176 & 47 & 16\\
Beam-off $\times$ Duty Ratio & 39 & 23 & 1 & 0\\
Beam-excess $\times$ Weight & 1716 & 153 $\times$ 1.48 & 46 & 16 $\times$
1.65\\ \hline
Total beam-excess & \multicolumn{2}{c}{1942 $\pm$ 46} \vline & \multicolumn{2}{c}{72.4
$\pm$ 9.5} \vline \\ 
\end{tabular}
\end{table}

\begin{table}
\caption{Beam-excess events, background, efficiency, neutrino
flux and flux-averaged cross section for the exclusive reaction
$^{12}$C($\nu_\mu$, $\mu^-$)$^{12}N_{g.s.}$}
\label{IV}
\begin{tabular}{lr}
Corrected beam excess events & 72.4 $\pm$ 9.5\\
$\bar{\nu}_\mu + ^{12}$C $\rightarrow \mu^+ + ^{12}$B$_{g.s.}$ background &
2.0 $\pm$ 0.4\\
accidental e$^+$ background & 13.6 $\pm$ 1.4\\ \hline
$\nu_\mu + ^{12}$C $\rightarrow \mu^- + ^{12}$N$_{g.s.}$ & 56.8 $\pm$ 9.6\\
``high energy $\mu$'' efficiency & 0.178 $\pm$ 0.013\\
$\nu_\mu$ flux (E$_\nu > 123.1$ MeV) & $1.33 \times 10^{12}$ cm$^{-2}$\\
$<\sigma >$ measured & $(6.6\pm 1.0\pm 1.0)\times 10^{-41}$ cm$^2$\\
$<\sigma >$ theory & $6.4 \times 10^{-41}$ cm$^2$\\
\end{tabular}
\end{table}

\begin{table}
\caption{Beam excess events, background, efficiency, neutrino flux and flux
averaged cross section for the inclusive reaction $^{12}$C($\nu_\mu$,
$\mu$)X.}
\label{V}
\begin{tabular}{ll}
Corrected Beam-excess events & 1942 $\pm$ 46\\
$\bar{\nu}_\mu +$ p $\rightarrow \mu^+ +$ n background & 140 $\pm$ 22\\
$\bar{\nu}_\mu + ^{12}$C $\rightarrow \mu^+ +$ X background & 46 $\pm$ 23\\
$\nu_\mu + ^{13}$C $\rightarrow \mu^- +$ X background & 18 $\pm$ 9\\ \hline
$\nu_\mu + ^{12}$C $\rightarrow \mu^- +$ X & 1738 $\pm$ 56\\
``high energy $\mu$'' efficiency & 0.313 $\pm$ 0.020\\
$\nu_\mu$ Flux (E$_\nu >$123.1 MeV) & 1.33 $\times 10^{12}$ cm$^{2}$\\
$<\sigma >$ measured & (11.2$\pm 0.3\pm 1.8)10^{-40}$ cm$^{2}$\\
$<\sigma >$ CRPA model& $20.5 \times 10^{-40}$ cm$^{2}$\\
\end{tabular}
\end{table}

\begin{table}
\caption{The expected and observed numbers of events with associated
neutrons and the calculated background from $\bar{\nu}_\mu$ reactions.}
\label{VI}
\begin{tabular}{|l|r|r|r|} 
& Events from & Fraction & Events with\\
Source & Table V & with neutron & neutron\\ \hline
Beam Excess & 1942 & (10.8$\pm$1.8)\% & 210 $\pm$ 35\\
$\bar{\nu}_\mu$ p $\rightarrow \mu^+$ n & 140 & 100\% & 140$\pm$22\\
$\bar{\nu}_\mu$ C $\rightarrow \mu^+$ nX & 46 & 79\% & 36$\pm$18\\ \hline
$\nu_\mu$C $\rightarrow \mu^-$X & 1756 & 1.9$\pm$2.6\% & 34$\pm$45\ 
\end{tabular}
\end{table}

\clearpage
\noindent

\begin{figure}
\centerline{\psfig{figure=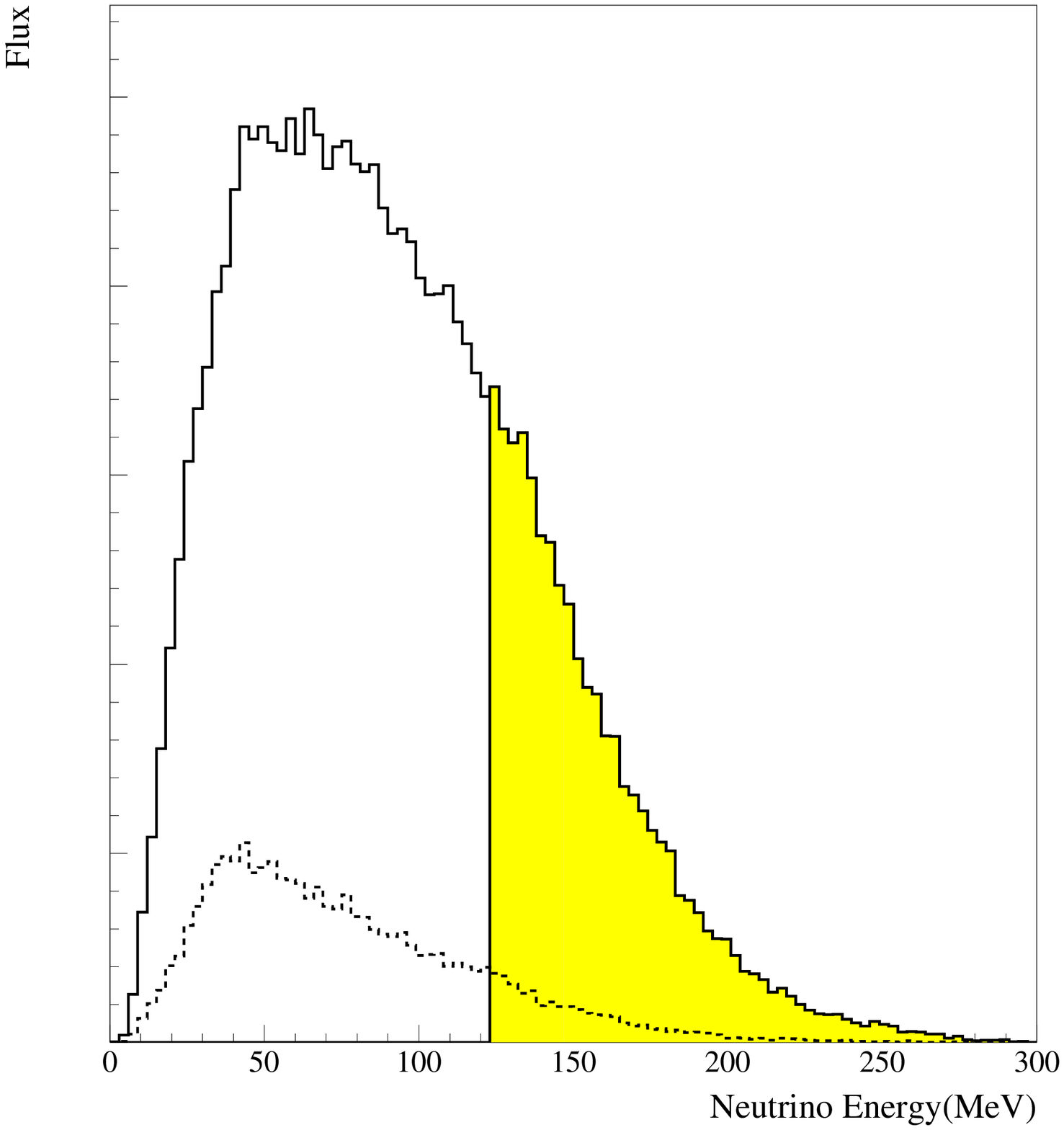,width=7.5in,silent=}}
\caption{       The solid line shows the flux shape of $\nu_\mu$ 
      from $\pi^+$ decay-in-flight.
      The region above muon production threshold is shaded. The dashed line
      shows the $\bar{\nu}_\mu$ flux from $\pi^-$ decay-in-flight for the
      same integrated proton beam. }
\label{Fig. 1}
\end{figure}

\begin{figure}
\centerline{\psfig{figure=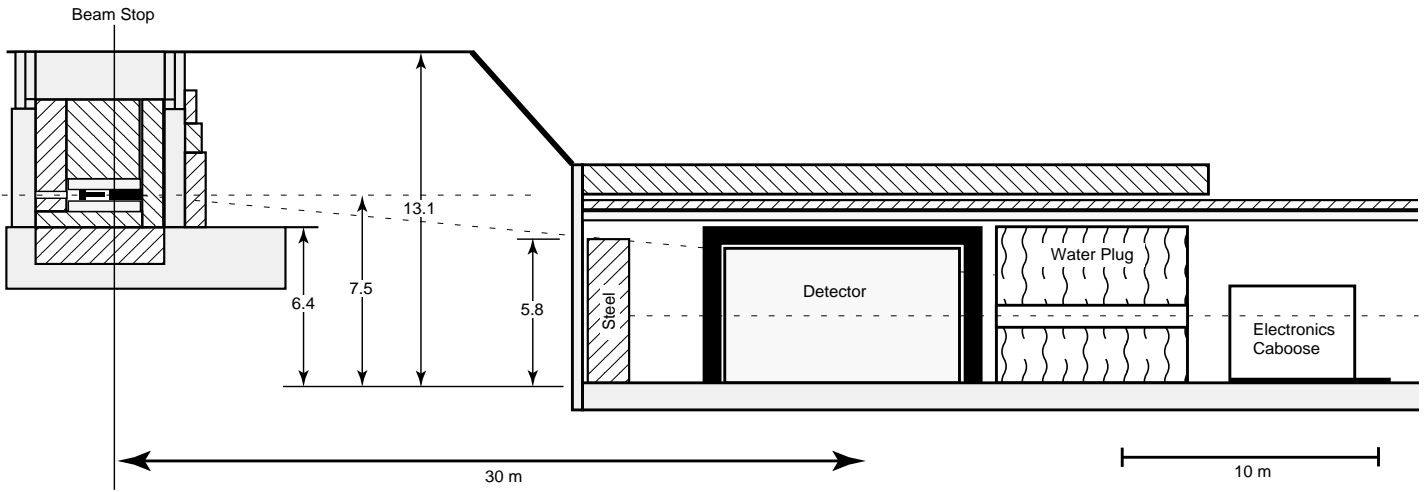,width=7.5in,silent=}}
\caption{The detector enclosure and target area configuration, elevation view.}
\label{Fig. 2}
\end{figure}

\begin{figure}
\centerline{\psfig{figure=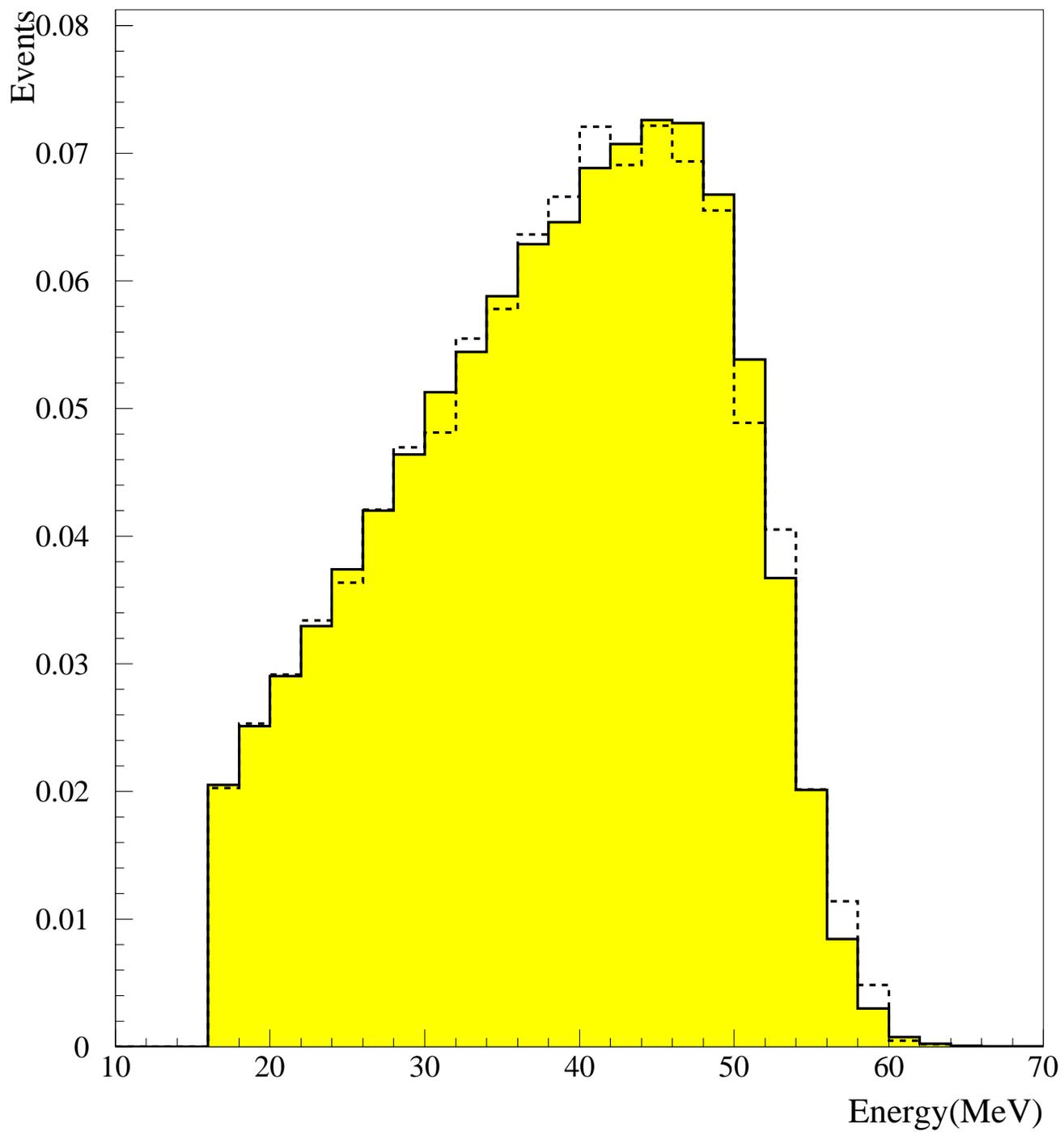,width=7.5in,silent=}}
\caption{The energy distribution for Michel electrons from data (solid) and
simulation (dashed).}
\label{Fig. 3}
\end{figure}

\begin{figure}
\centerline{\psfig{figure=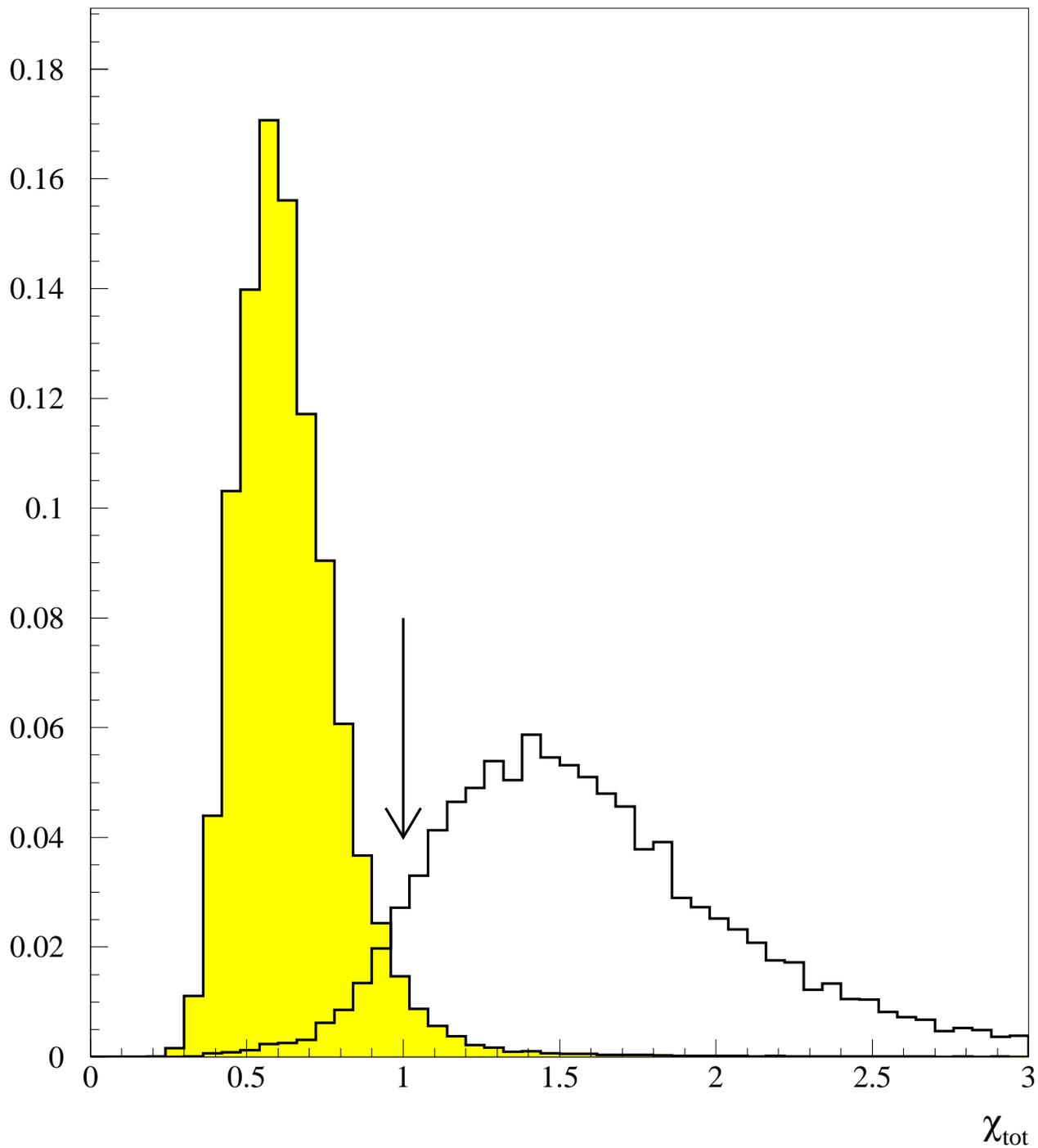,width=7.5in,silent=}}
\caption{Particle ID parameter for ``electrons'' (shaded) and ``neutrons''.}
\label{Fig. 4}
\end{figure}

\begin{figure}
\centerline{\psfig{figure=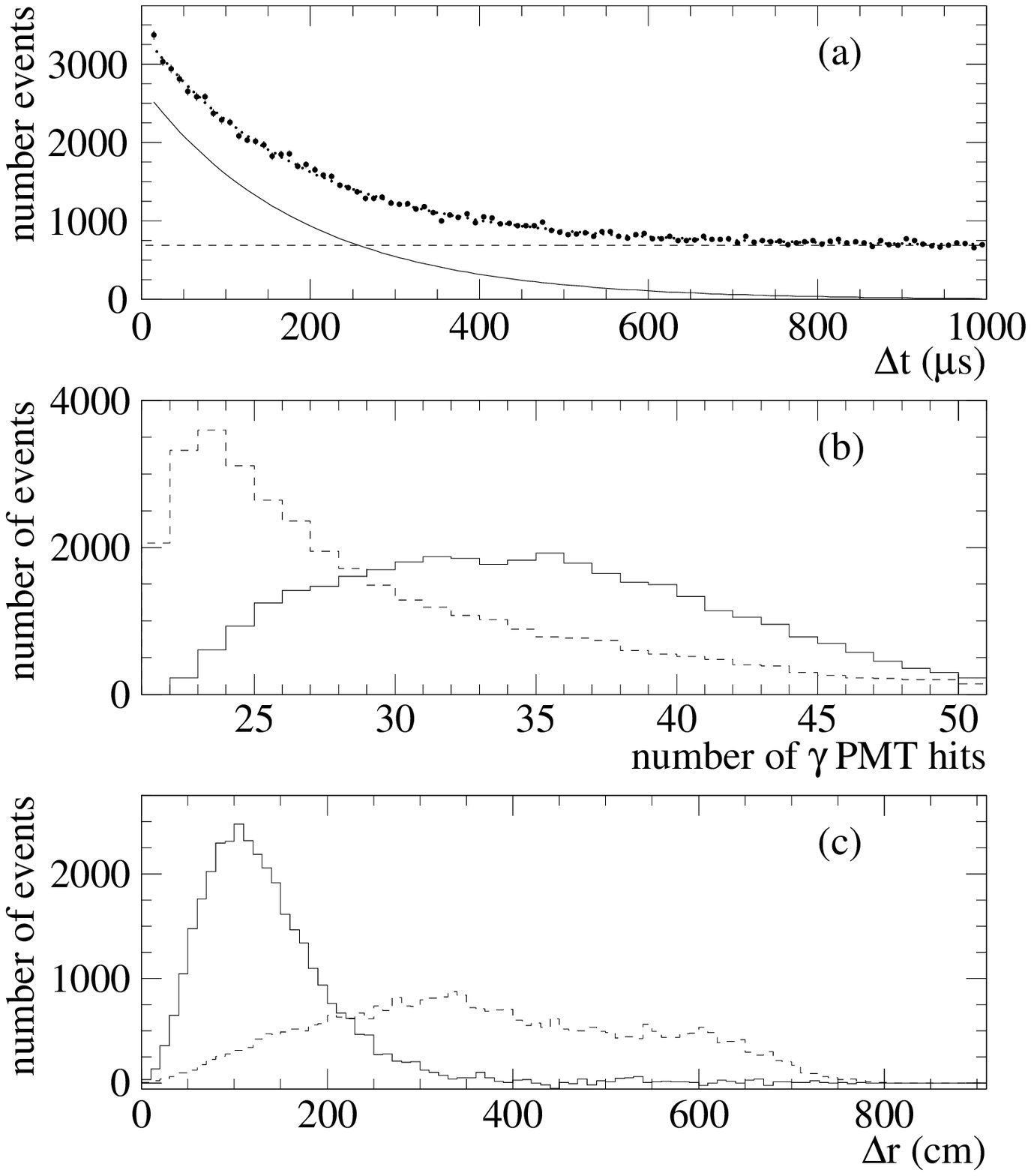,width=7.5in,silent=}}
\caption{Distributions obtained from cosmic-ray neutron data for $\gamma$'s
that are correlated (solid) or uncorrelated (dashed) with the primary event:
(a) the time between the photon and primary event; (b) the number of photon
PMT hits; and (c) the distance between the photon and primary event.  The
raw data points are also shown in (a).}
\label{Fig. 5}
\end{figure}

\begin{figure}
\centerline{\psfig{figure=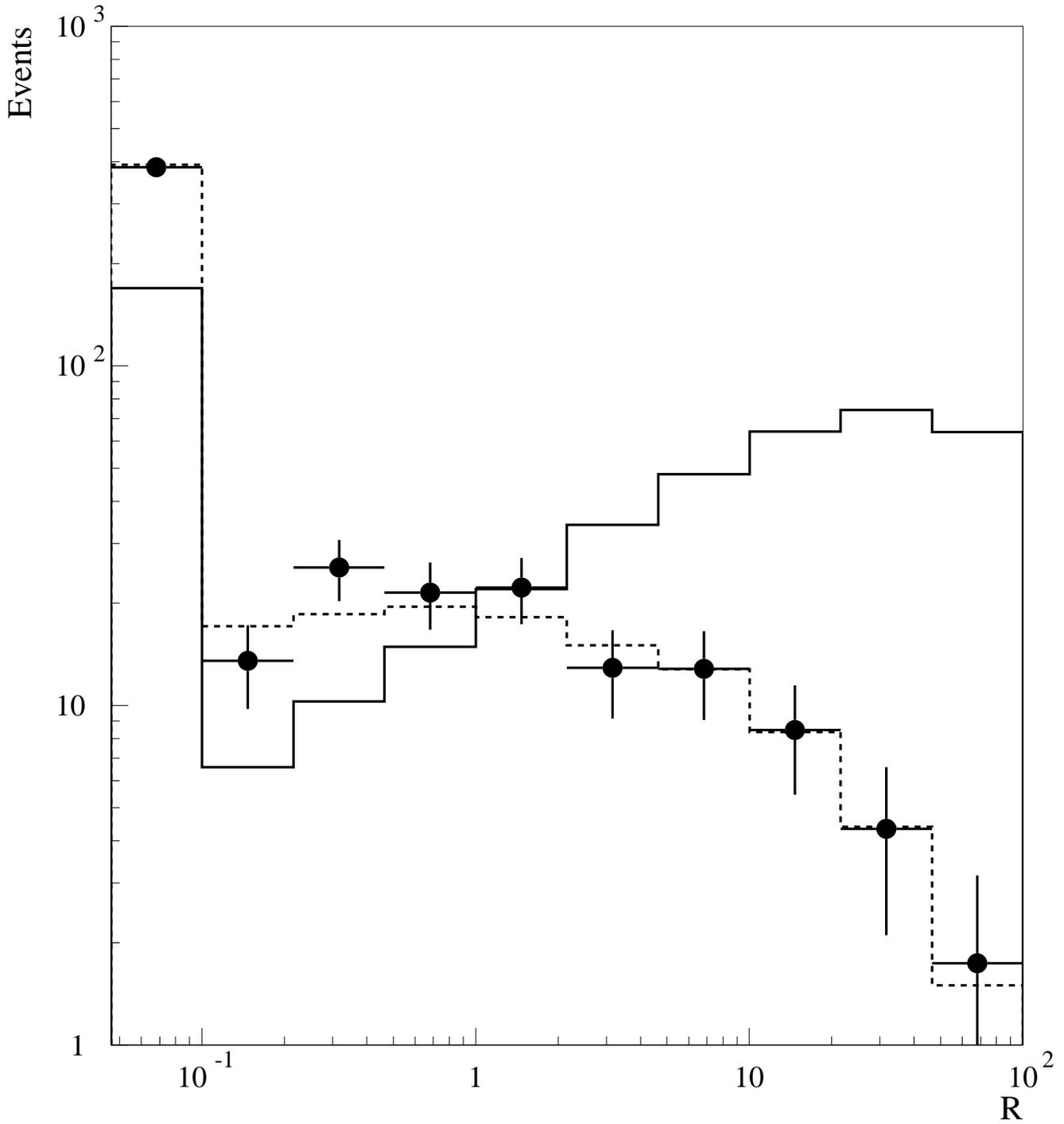,width=7.5in,silent=}}
\caption{The measured R distribution for events with the $\gamma$ correlated
(solid) or uncorrelated (dashed) with the primary event.  Also shown is the
observed R distribution for a neutrino process with no correlated $\gamma$'s,
$^{12}$C($\nu_e$,e$^-$)$^{12}$N$_{g.s.}$.}
\label{Fig. 6}
\end{figure}

\begin{figure}
\centerline{\psfig{figure=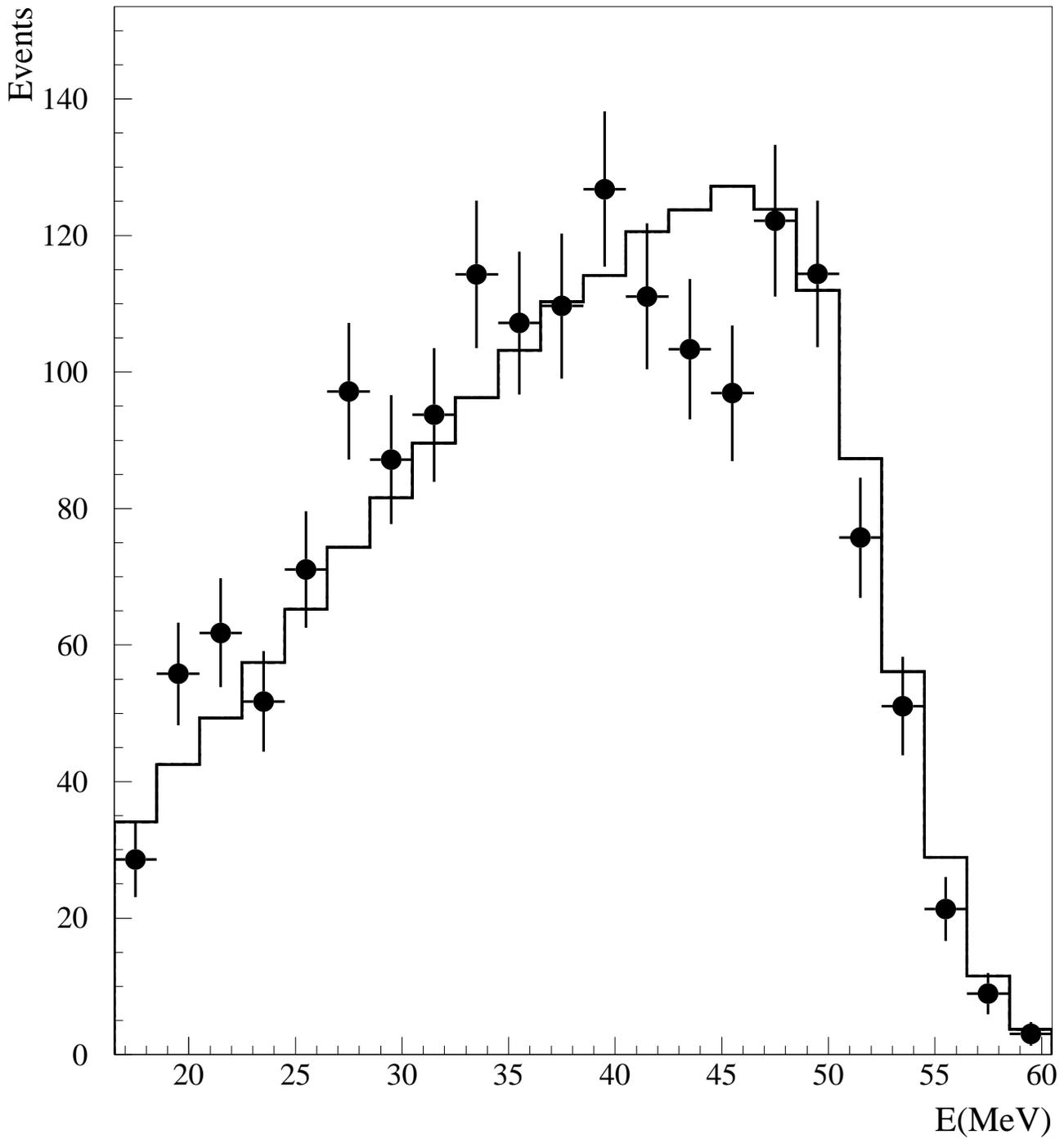,width=7.5in,silent=}}
\caption{The observed energy spectra of electrons from the decays of muons
for the inclusive neutrino sample (points) and for the stopping cosmic ray muon
sample(solid line).}
\label{Fig. 7}
\end{figure}

\begin{figure}
\centerline{\psfig{figure=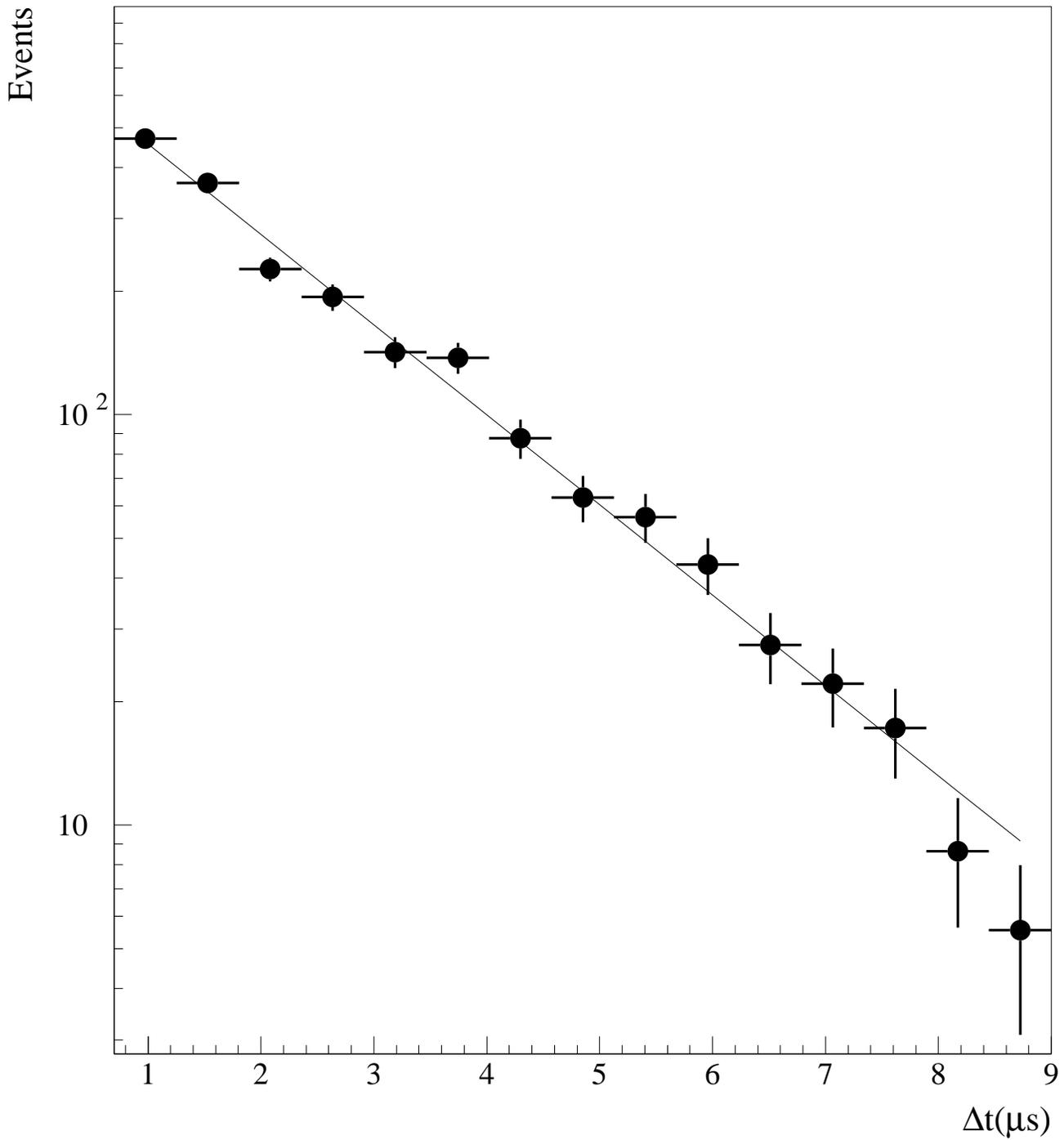,width=7.5in,silent=}}
\caption{The distribution of time $\Delta$t$_{\mu e}$ between the
$\mu^-$ and the decay e$^-$ in
the inclusive sample, $^{12}$C($\nu_\mu$,$\mu^-$)X. The best fit(solid line) curve 
corresponds to a lifetime of $1.98 \pm 0.06 \mu$s. }
\label{Fig. 8}
\end{figure}

\begin{figure}
\centerline{\psfig{figure=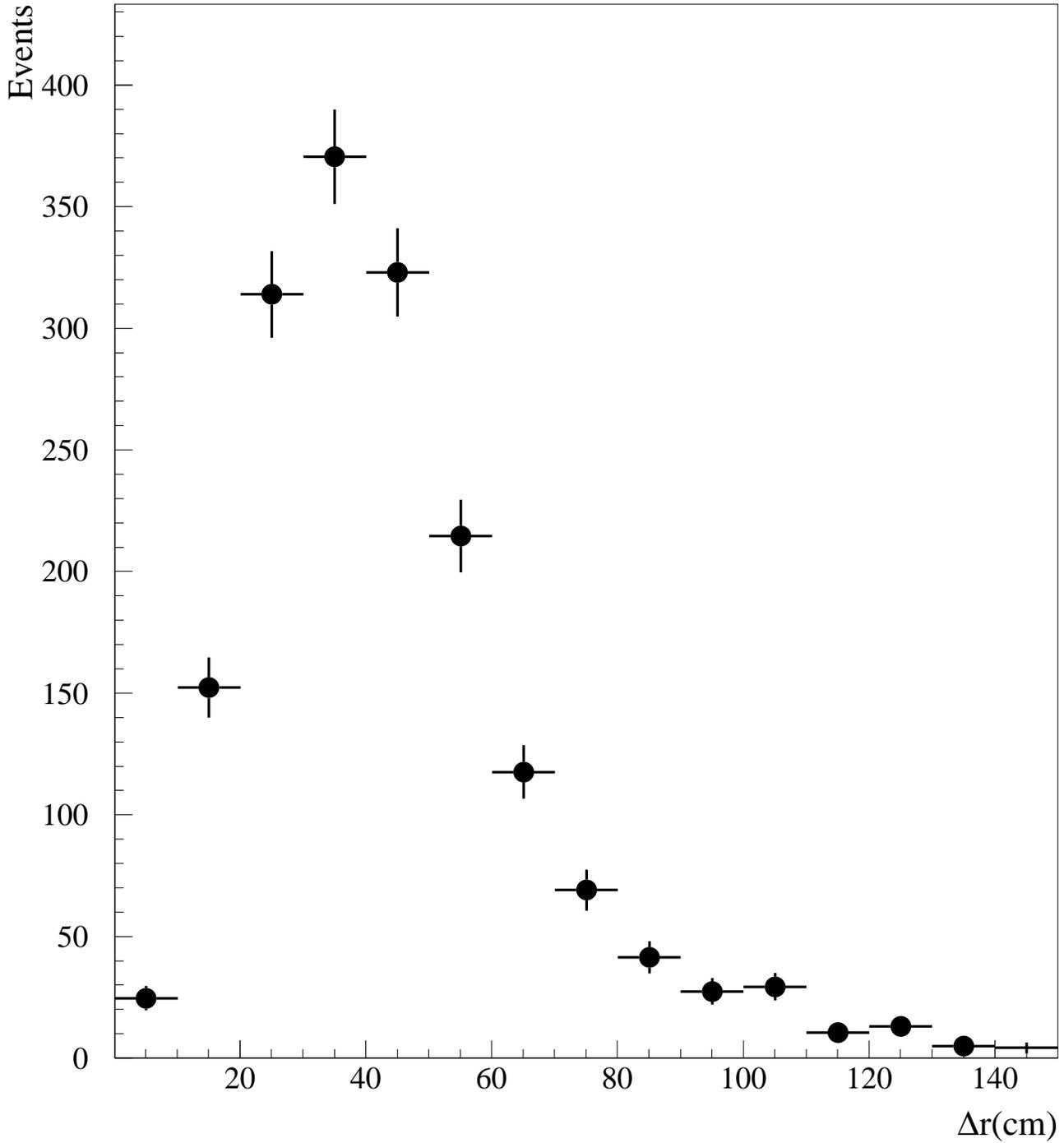,width=7.5in,silent=}}
\caption{The distribution of the distance between the reconstructed position
of the
$\mu^-$ and the e$^-$ in the beam-excess 
inclusive sample, $^{12}$C($\nu_\mu$,$\mu^-$)X.}
\label{Fig. 9}
\end{figure}

\begin{figure}
\centerline{\psfig{figure=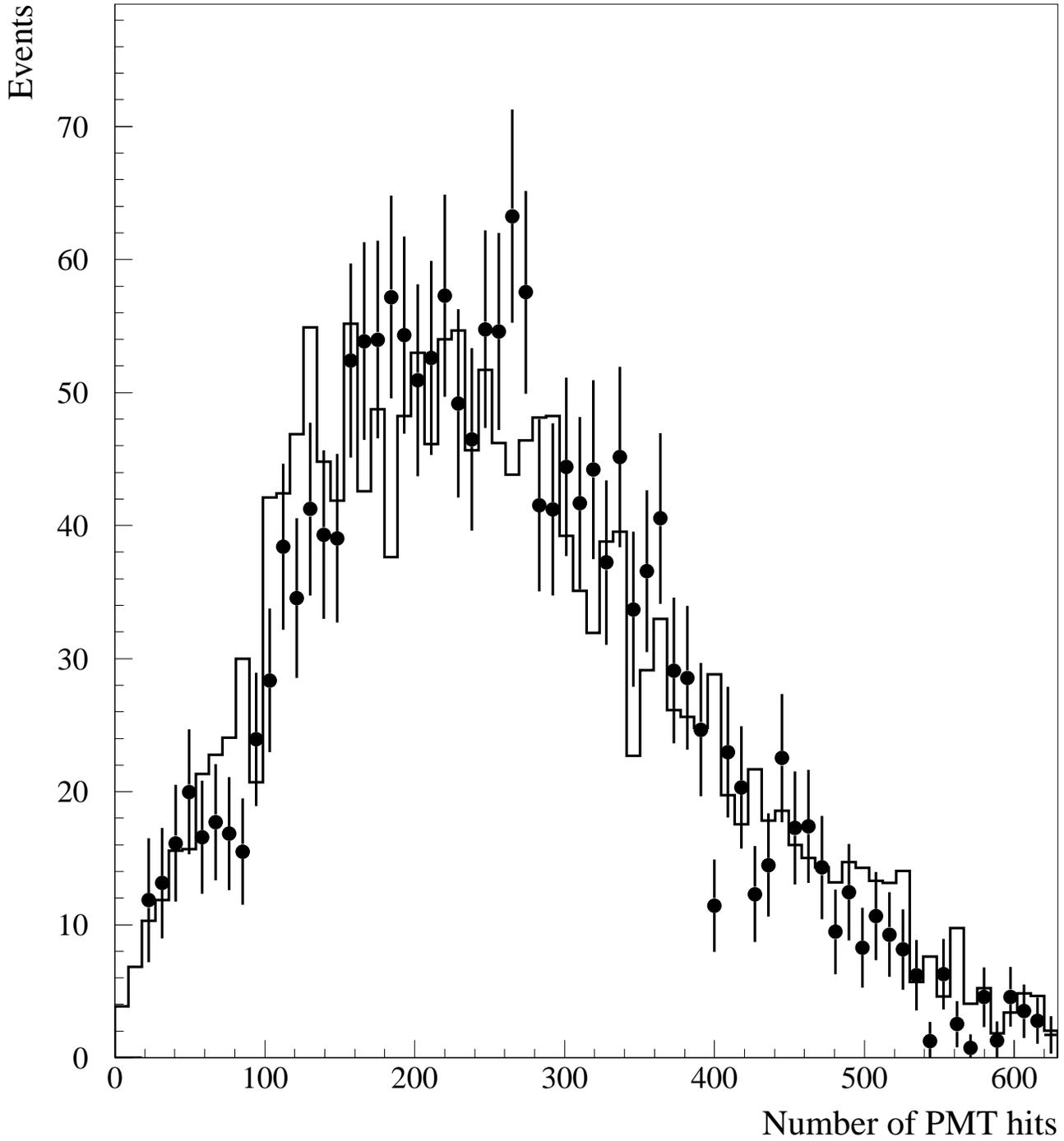,width=7.5in,silent=}}
\caption{The observed PMT hit distribution for the decay-in-flight sample
(including $\nu_{\mu} C \rightarrow \mu^- X$, $\bar \nu_{\mu} C 
\rightarrow \mu^+ X$,
and $\bar \nu_{\mu} p \rightarrow \mu^+ n$).
The solid histogram is the prediction from the
Monte Carlo simulation, normalized to the data.}
\label{Fig. 10}
\end{figure}

\begin{figure}
\centerline{\psfig{figure=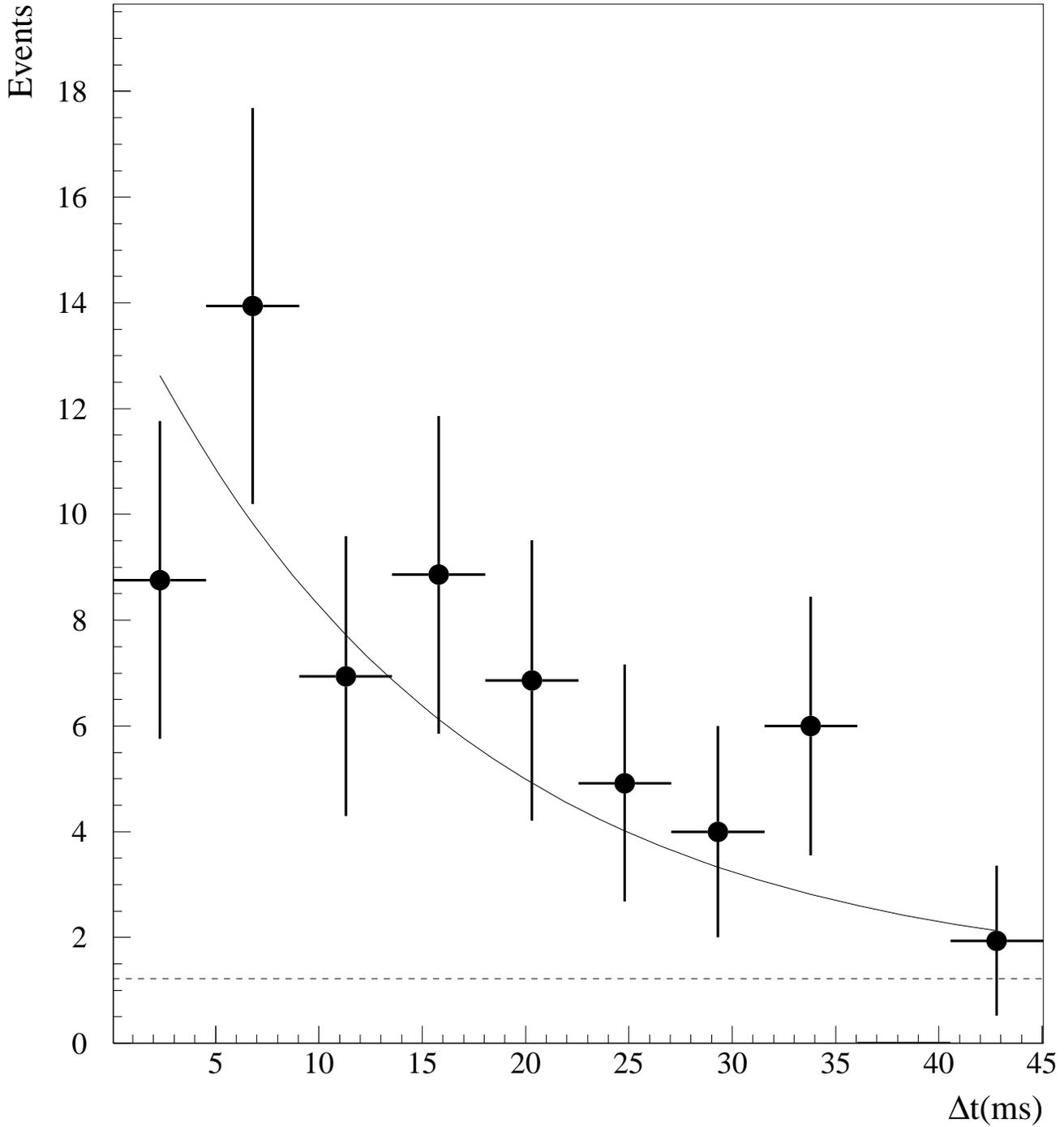,width=7.5in,silent=}}
\caption{The distribution of time between the e$^-$ and e$^+$ for beam-excess
events
in the $^{12}$C($\nu_\mu$,$\mu^-$)$^{12}$N$_{g.s.}$ sample.  The expected
distribution is shown with the solid line.  The calculated accidental
contribution is shown by the dashed line.}
\label{Fig. 11}
\end{figure}

\begin{figure}
\centerline{\psfig{figure=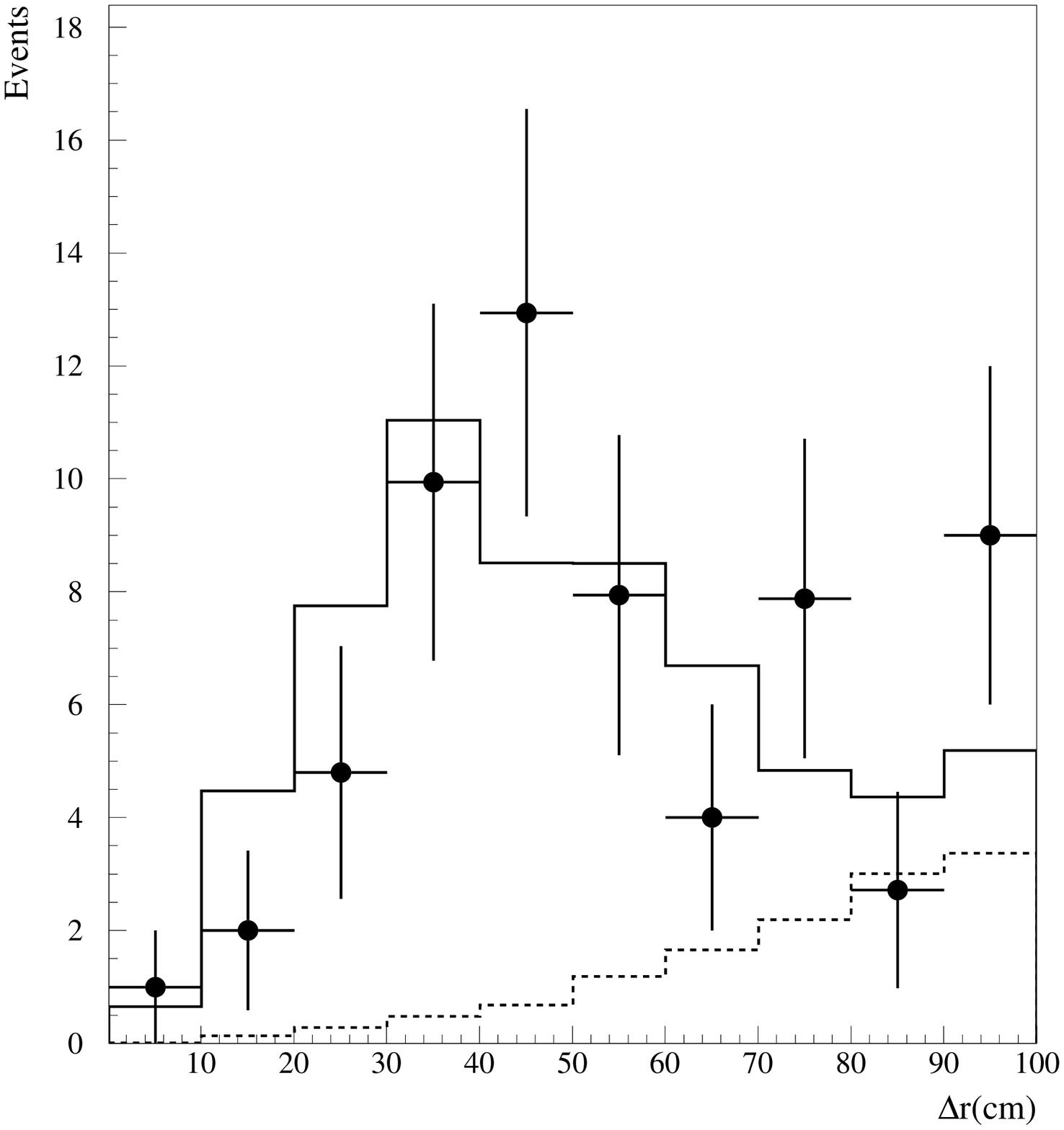,width=7.5in,silent=}}
\caption{The distribution of the distance between the reconstructed positions of
e$^-$ and e$^+$ for beam-excess events in the  $^{12}$C($\nu_\mu$,$\mu^-$)$^{12}$N$_{g.s.}$ sample.
The dashed line shows the calculated
       accidental contribution.  
The solid line shows the expected shape
       (including the accidental contribution) obtained with a large
sample of events from the reaction $^{12}$C($\nu_e$,$e^-$)$^{12}$N$_{g.s.}$.  }
\label{Fig. 12}
\end{figure}

\begin{figure}
\centerline{\psfig{figure=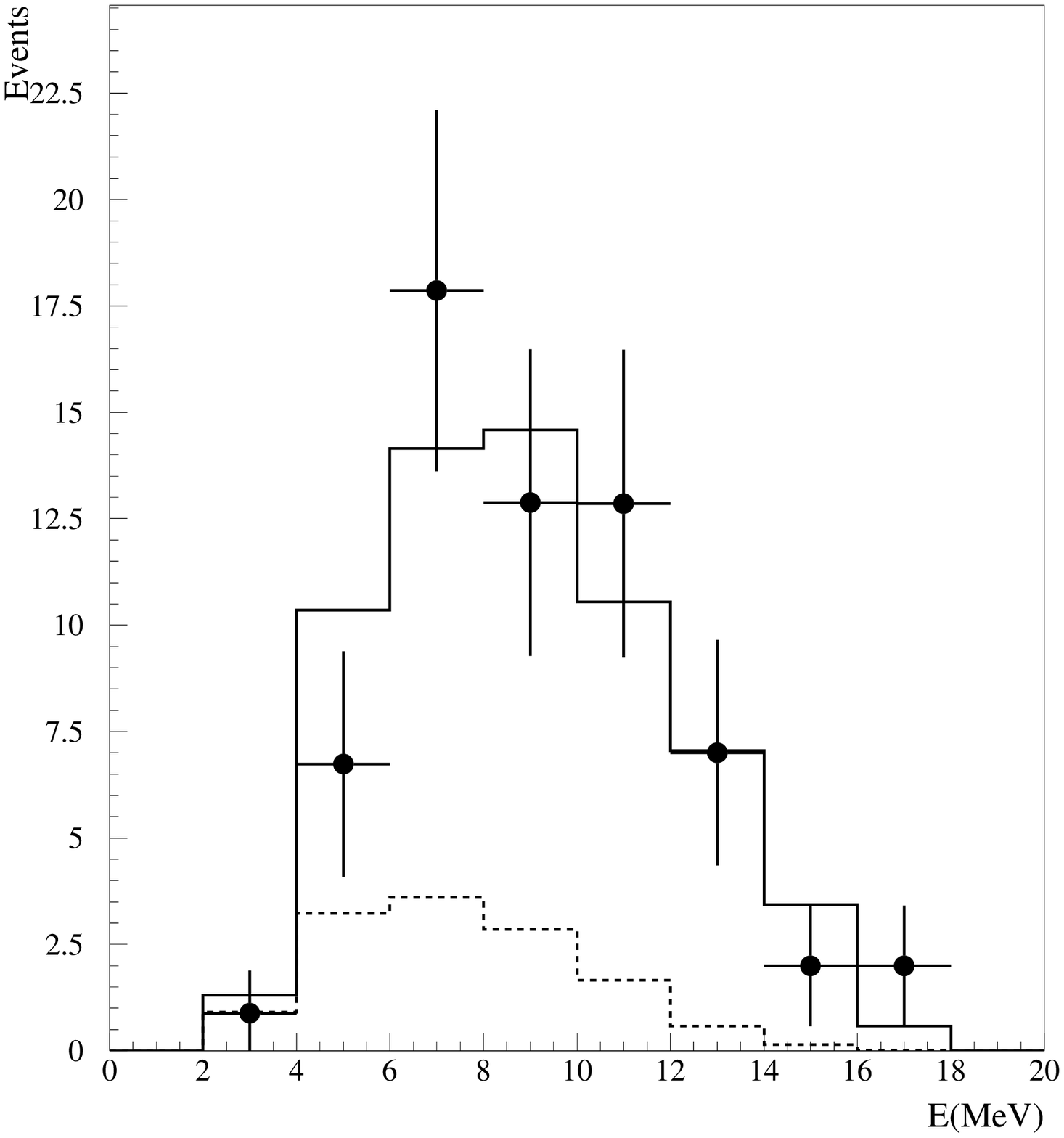,width=7.5in,silent=}}
\caption{The observed and expected (solid line) e$^+$ energy distribution for
beam-excess events in the $^{12}$C($\nu_\mu$,$\mu^-$)$^{12}$N$_{g.s.}$
sample. The expected distribution was obtained with a large
sample of events from the reaction $^{12}$C($\nu_e$,$e^-$)$^{12}$N$_{g.s.}$.
The dashed line shows the expected contribution from background sources.}
\label{Fig. 13}
\end{figure}

\begin{figure}
\centerline{\psfig{figure=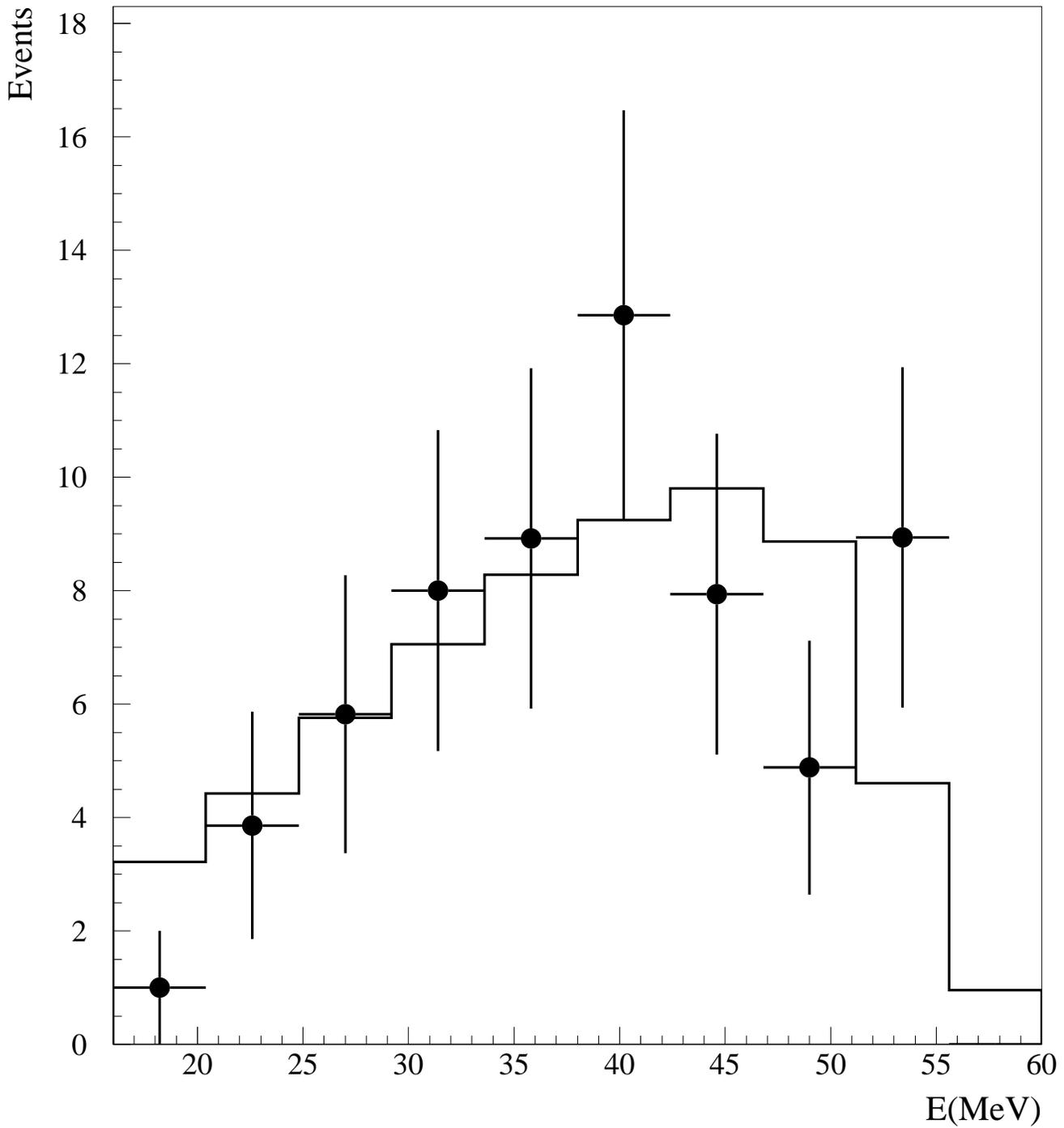,width=7.5in,silent=}}
\caption{The observed and expected (solid line) e$^-$ energy distribution for
beam-excess events in the $^{12}$C($\nu_\mu$,$\mu^-$)$^{12}$N$_{g.s.}$
sample. The expected distribution was obtained from a large
sample of stopping cosmic ray events. }
\label{Fig. 14}
\end{figure}

\begin{figure}
\centerline{\psfig{figure=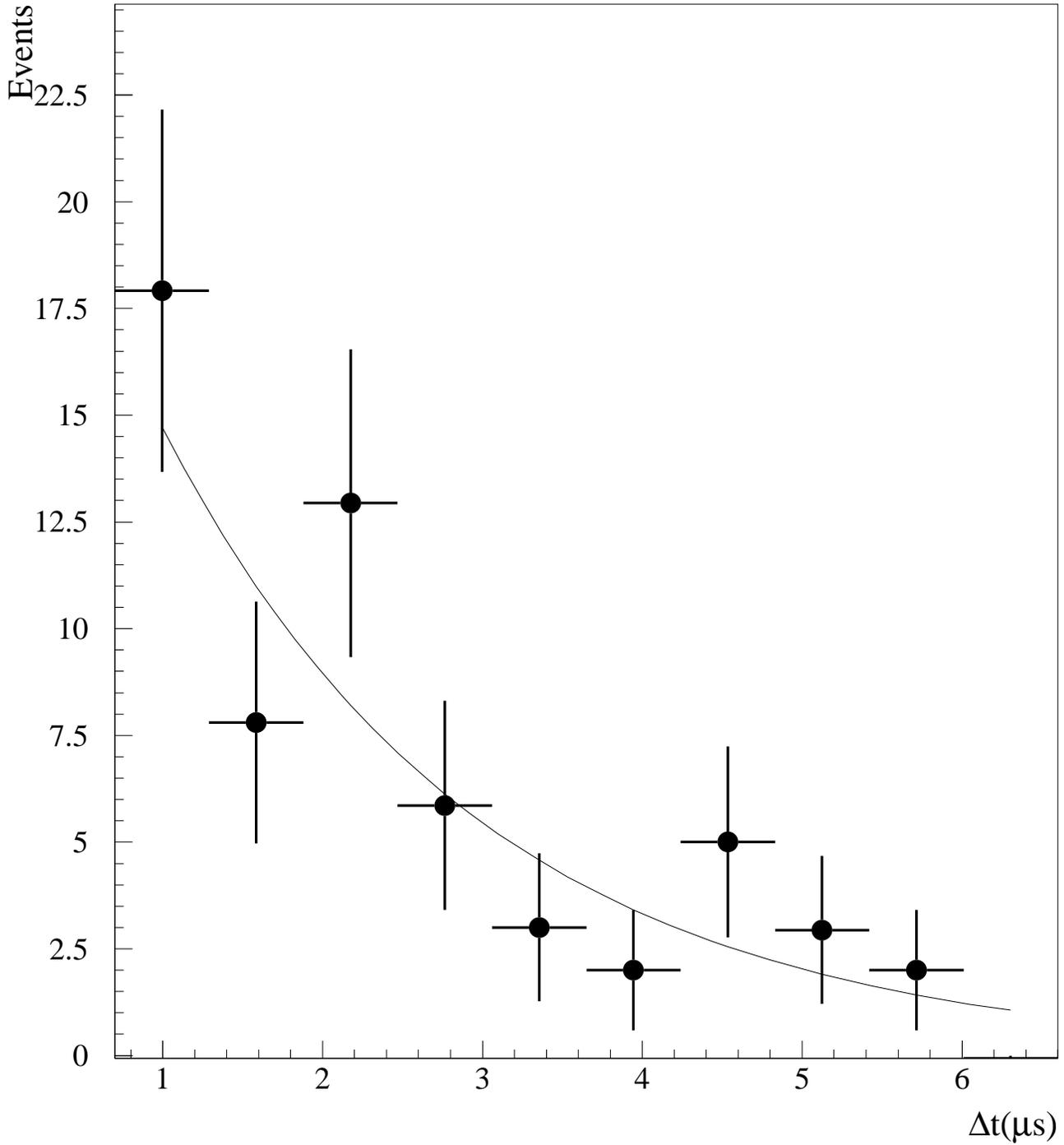,width=7.5in,silent=}}
\caption{The obtained and expected (solid line) distribution of time between the $\mu^-$ and the subsequent decay
e$^-$ for beam-excess events in the $^{12}$C($\nu_\mu$,$\mu^-$)$^{12}$N$_{g.s.}$
sample.}
\label{Fig. 15}
\end{figure}

\begin{figure}
\centerline{\psfig{figure=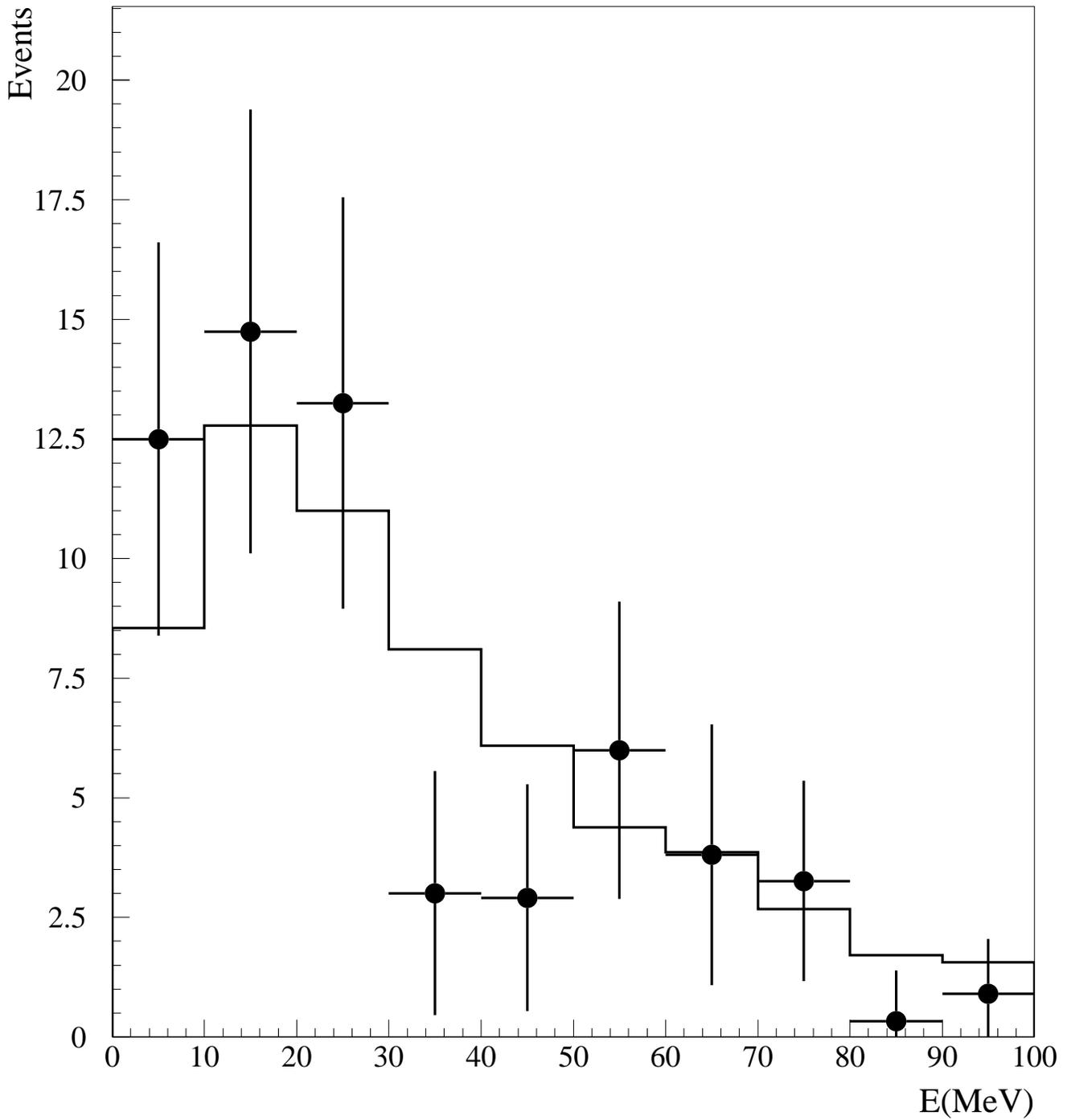,width=7.5in,silent=}}
\caption{The observed and expected (solid line) $\mu^-$ kinetic 
energy distribution for
beam excess-events in the $^{12}$C($\nu_\mu$,$\mu^-$)$^{12}$N$_{g.s.}$
sample.}
\label{Fig. 16}
\end{figure}

\begin{figure}
\centerline{\psfig{figure=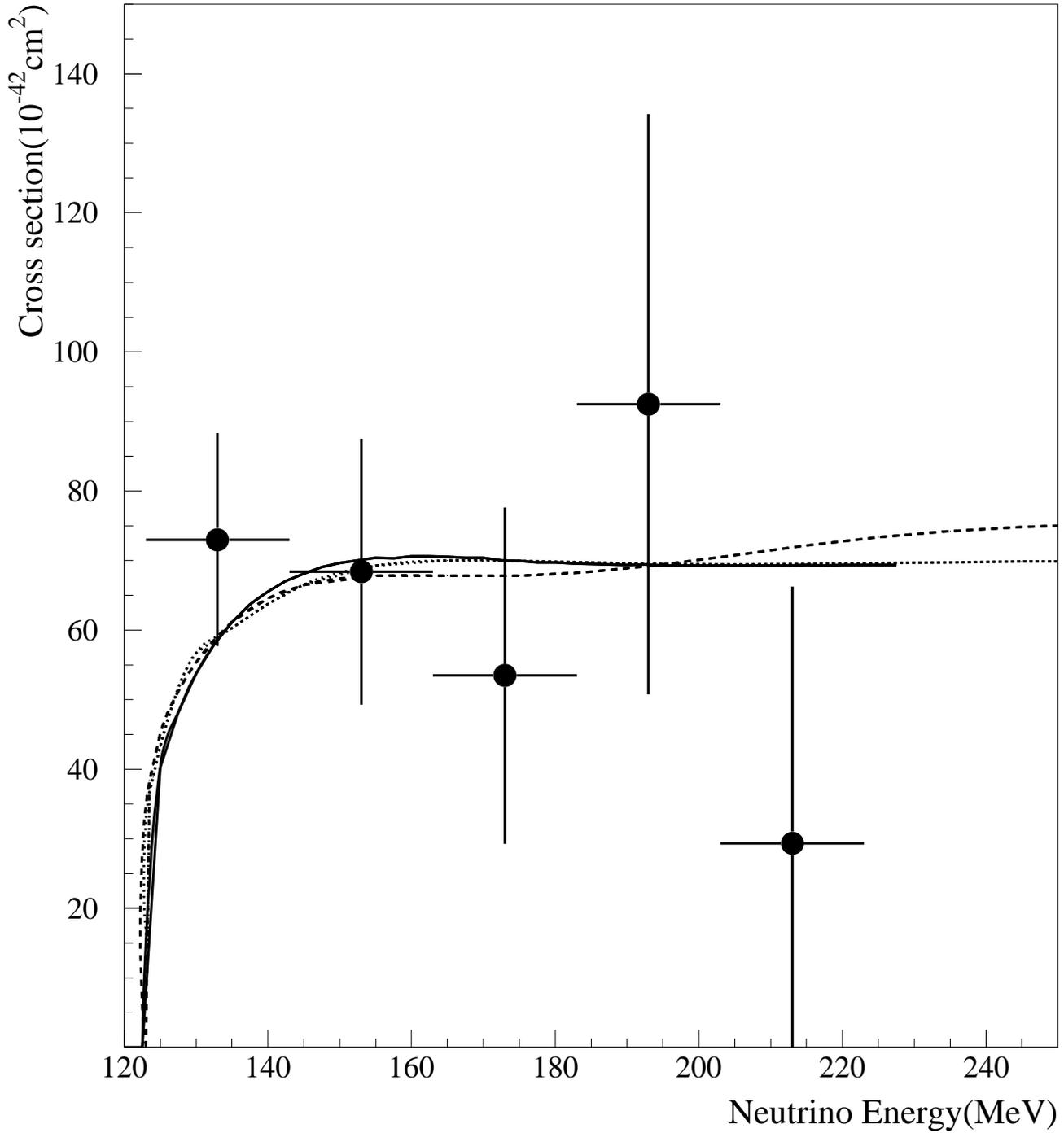,width=7.5in,silent=}}
\caption{The measured cross section for the
process $^{12}$C($\nu_\mu$,$\mu^-$)$^{12}$N$_{g.s.}$ compared with three
theoretical calculations obtained from Ref. 20.}
\label{Fig. 17}
\end{figure}

\begin{figure}
\centerline{\psfig{figure=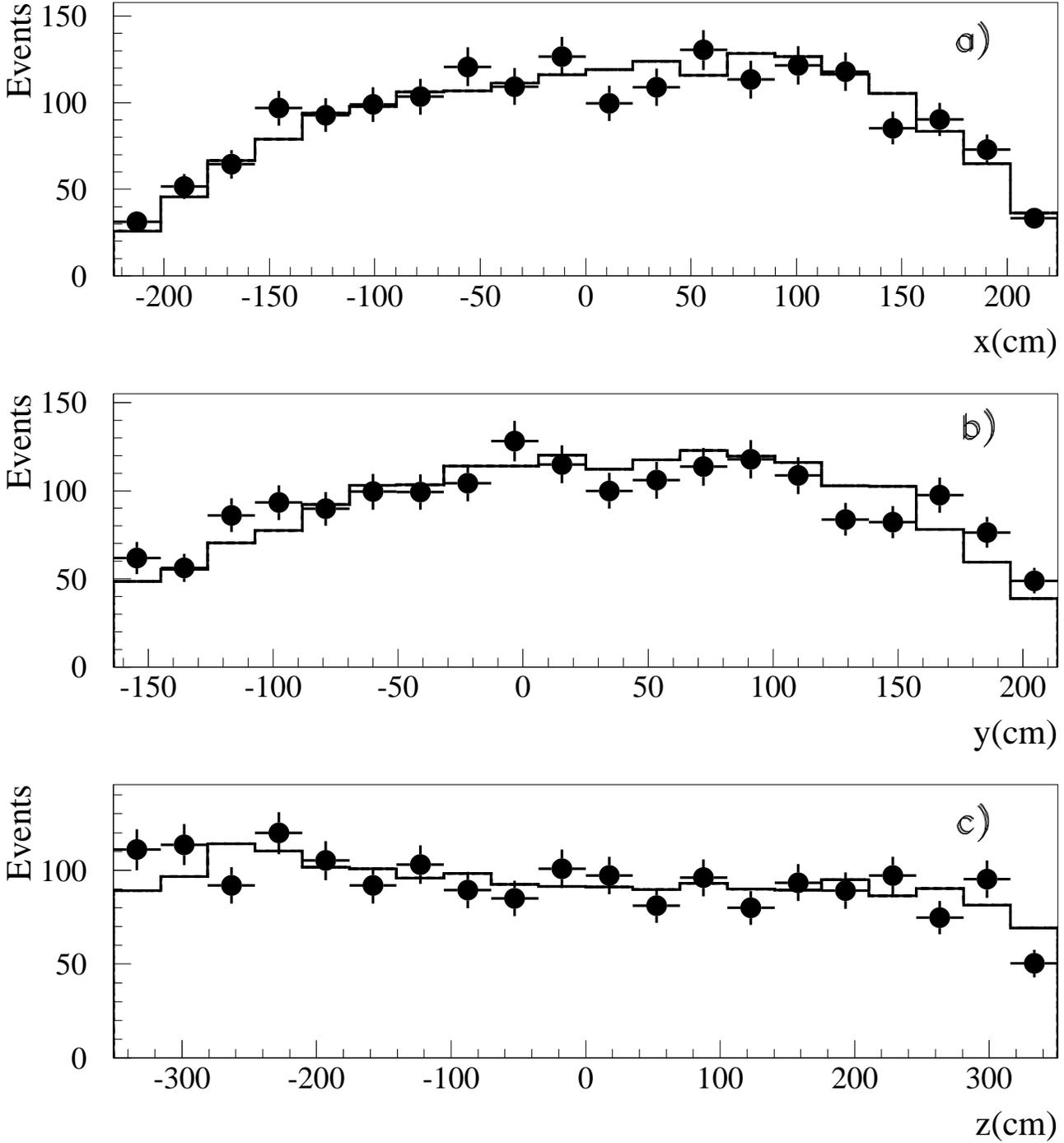,width=7.5in,silent=}}
\caption{The observed spatial distribution of beam-excess electrons compared with
the expected (solid line) distribution for the
process $^{12}$C($\nu_\mu$,$\mu^-$)$^{12}$X}
\label{Fig. 18}
\end{figure}

\begin{figure}
\centerline{\psfig{figure=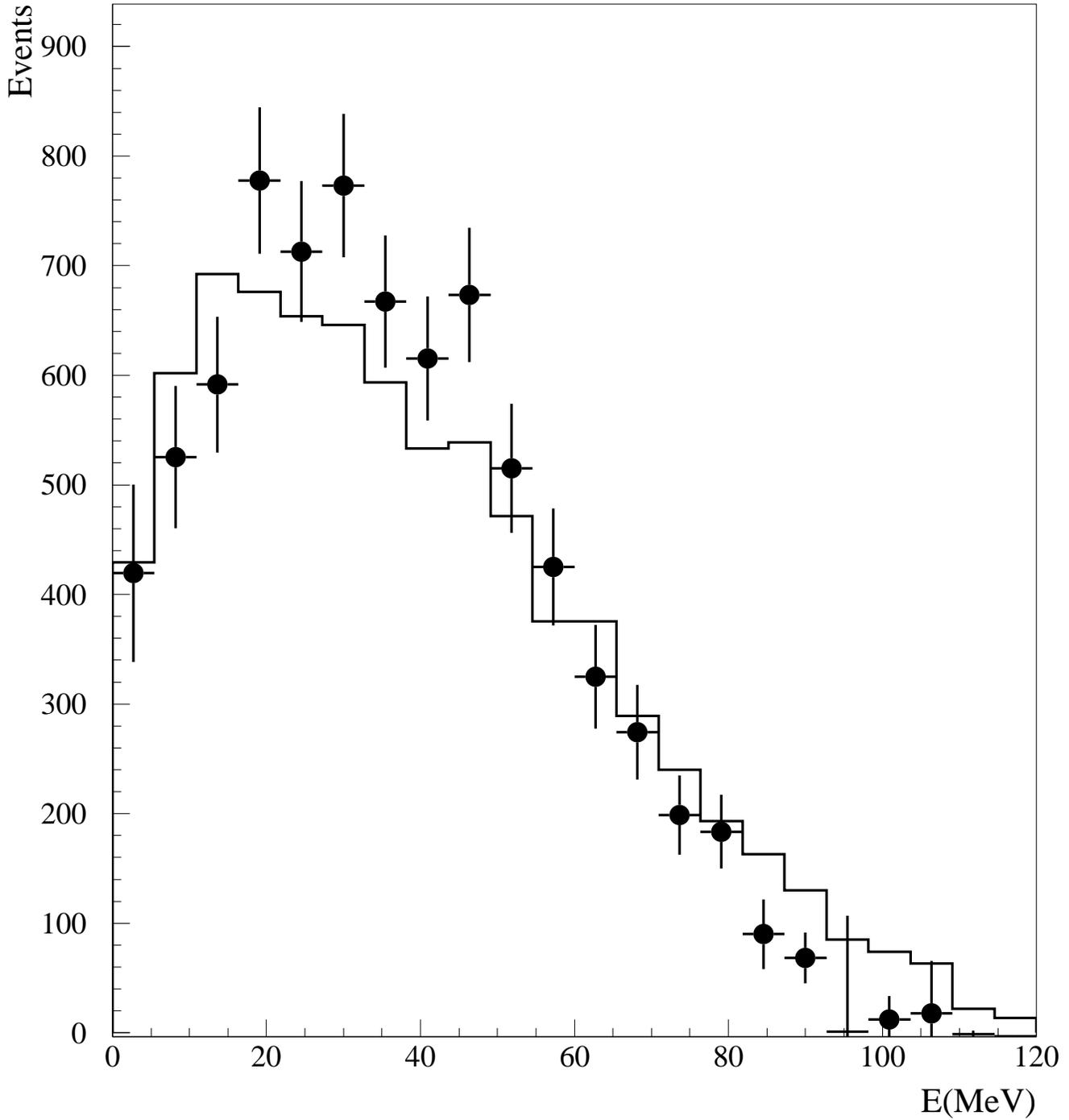,width=7.5in,silent=}}
\caption{The observed and expected distribution of the muon
kinetic energy, E$_\mu$, for the inclusive decay-in-flight sample.
The expected distribution has been normalized to the data.}
\label{Fig. 19}
\end{figure}

\begin{figure}
\centerline{\psfig{figure=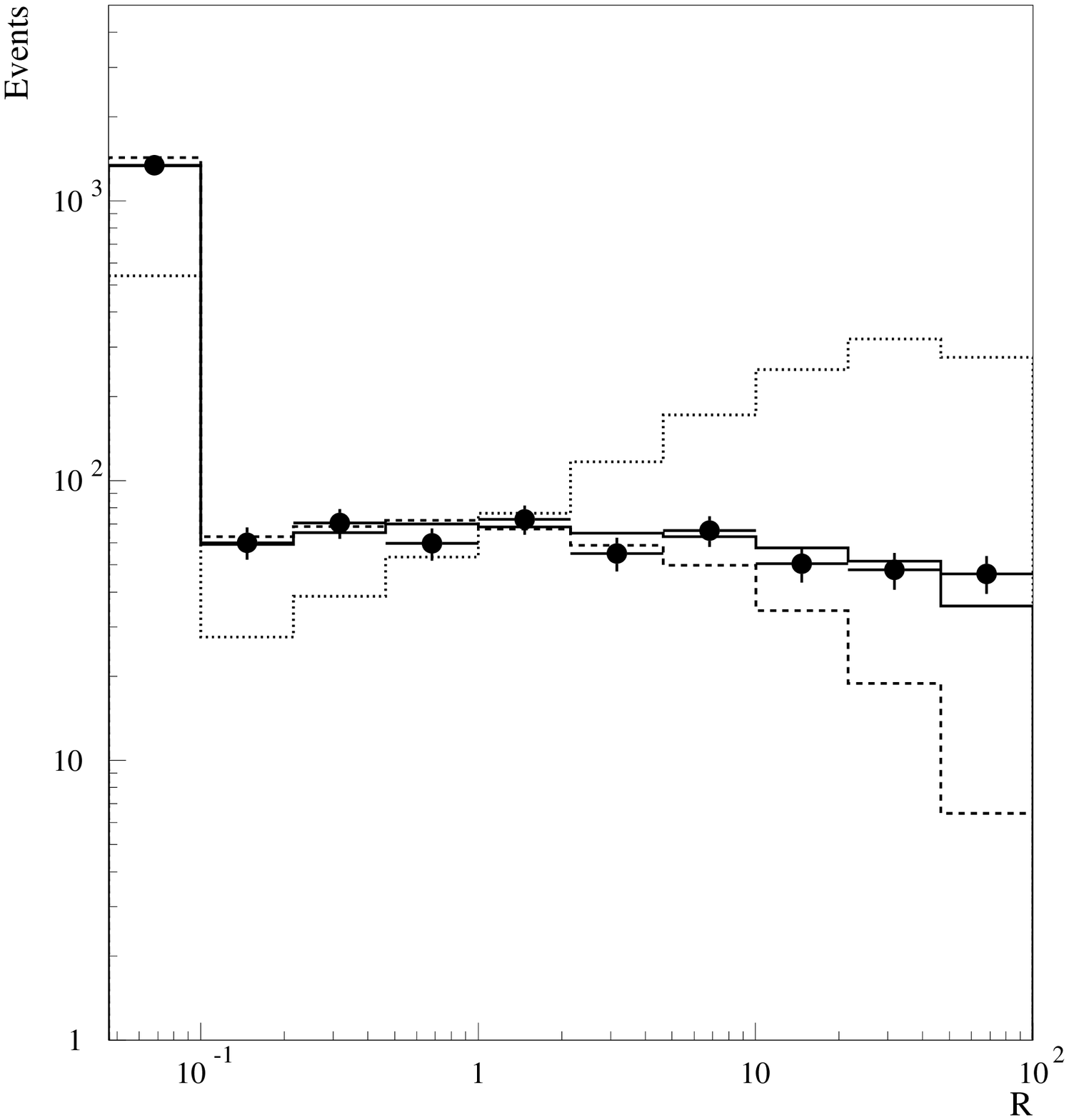,width=7.5in,silent=}}
\caption{The observed distribution of the $\gamma$ likelihood ratio R for
the inclusive decay-in-flight sample.  Shown for comparison are the
correlated distribution (dotted line), the uncorrelated distribution (dashed
line) and the best fit (solid line) to the data which has a (10.8$\pm$1.8)\%
correlated component.}
\label{Fig. 20}
\end{figure}

{\Large\bf REFERENCES}

\begin{enumerate}
\item D.A. Krakauer {\it et al.}, Phys. Rev. C {\bf 45,} 2450 (1992).
\item B.E. Bodmann {\it et al.}, Phys. Lett. B {\bf 332,} 251 (1994). 
\item C. Athanassopoulos {\it et al.}, Phys. Rev. C {\bf 55,} 2078 (1997).
\item M. Fukugita {\it et al.}, Phys. Lett. B {\bf 212,} 139 (1988).
\item E. Kolbe, K. Langanke and S. Krewald, Phys. Rev. C {\bf 49,} 1122 (1994).
\item M. Albert {\it et al.}, Phys. Rev. C {\bf 51,} 1065 (1995).
\item H.J. Kim {\it et. al} Proceedings of the 14th International 
Conference on Particles and Nuclei (PANIC 96), 583 (1996).
\item E. Kolbe {\it et al.}, Phys. Rev. C {\bf 52,} 3437 (1995).
\item A. C. Hayes,  paper presented at the Joint APS/AAPT meeting,
April, 18-21, 1997
\item D.D. Koetke {\it et al.}, Phys. Rev. C {\bf 46,} 2554 (1992).
\item C. Athanassopoulos {\it et al.}, Phys. Rev. C {\bf 54,} 2685 (1996).
\item C. Athanassopoulos {\it et al.}, submitted to Nucl. Instrum.
Methods.
\item S. Willis {\it et al.}, Phys. Rev. Lett. {\bf 44,} 522 (1980).
\item S.J. Freedman {\it et al.}, Phys. Rev. D {\bf 47,} 811 (1993).
\item R.C. Allen, {\it et al.}, Nucl. Instrum. Methods A {\bf 284,} 347
(1989).
\item R.L. Burman, M.E. Potter and E.S. Smith, Nucl. Instrum. Methods A
{\bf 291,} 621 (1990).
\item R.A. Reeder {\it et al.}, Nucl. Instrum. Methods A {\bf 334,} 353
(1993).
\item J. Napolitano, {\it et al.}, Nucl. Instrum. Methods A {\bf 274,} 152
(1989).
\item K. McIlhany, {\it et al.}, Proceedings of the Conference on Computing
in High Energy Physics, April 1994, LBL Report 35822, 357 (1995).
\item J. Engel, E. Kolbe, K. Langanke and P. Vogel, Phys. 
Rev. C {\bf 54,} 2740 (1996).
\item R. Becker-Szendy {\it et al.} (IMB Collaboration), 
Phys. Rev. D {\bf 46,} 3270 (1992).
\item K. S. Hirata {\it et al.} (Kamiokande Collaboration), 
Phys. Lett. B {\bf 280} 146 (1992).

\end{enumerate}

\clearpage

\end{document}